\documentclass[aps,prd,preprint,nofootinbib,a4paper,11pt,superscriptaddress]{revtex4-1}% добавляется twoside для двусторонней печати

\usepackage[T1]{fontenc}
\usepackage[centertags]{amsmath}
\usepackage{amsfonts}
\usepackage{amssymb}
\usepackage{amsthm}
\usepackage[sort&compress]{natbib}% упорядочивает ссылки
\usepackage[hypertex]{hyperref}%hypertex

\DeclareMathOperator{\re}{Re}

\DeclareMathOperator{\Tr}{Tr}
\DeclareMathOperator{\Sp}{Sp}

\DeclareMathOperator{\sgn}{sgn}
\DeclareMathOperator{\Ai}{Ai}
\DeclareMathOperator{\Bi}{Bi}
\newcommand{\lan}{\langle}
\newcommand{\ran}{\rangle}

\newcommand{\e}{\varepsilon}
\newcommand{\vf}{\varphi}

\newcommand{\s}{\sigma}

\newcommand{\al}{\alpha}
\newcommand{\be}{\beta}
\newcommand{\ga}{\gamma}
\newcommand{\Ga}{\Gamma}
\newcommand{\de}{\delta}

\newcommand{\la}{\lambda}
\newcommand{\La}{\Lambda}

\newcommand{\N}{\mathbb{N}}

\newtheorem{thm}{Theorem}

\begin{document}
\setlength{\unitlength}{1pt}% устанавливает единицу длины в окружении picture

\title{High-temperature expansion of the grand thermodynamic potential  for scalar particles in crossed electromagnetic fields}

\author{I.S. Kalinichenko}
\email[E-mail:]{probustom@gmail.com}
\affiliation{Department of Physics, Tomsk State University, Tomsk 634050, Russia}

\author{P.O. Kazinski}
\email[E-mail:]{kpo@phys.tsu.ru}
\affiliation{Department of Physics, Tomsk State University, Tomsk 634050, Russia}

\date{\today}

\begin{abstract}

The problem of a scalar particle in a constant crossed electromagnetic field ($\mathbf{E}\perp\mathbf{H}$ and $|\mathbf{E}|=|\mathbf{H}|$) is considered. The high-temperature expansion of the one-loop grand thermodynamic potential and vacuum energy with account for non-perturbative corrections are derived. The contributions from particles and antiparticles are obtained separately. It is shown that the non-perturbative corrections depend on the boundary conditions but do not depend on the fields.

\end{abstract}

\maketitle

\section{Introduction}

Usually, the one-loop contribution to the effective action in constant electromagnetic fields including the nonzero electric component is obtained either by solving the Heisenberg equations for a quantum particle in this field \cite{Schwing.10}, or by analytic continuation of the result for purely magnetic fields together with the Poincare-invariance arguments \cite{HeisEul,LandLifQED}, or, at finite temperature, by summing the leading in derivatives contributions to the effective action that do not include the strength of the electromagnetic field \cite{ElmfSkag95}(for finite temperature approaches, see also \citep{Loewe,Ganguly,Shovkovy,DittGies}). As for the crossed electromagnetic fields, the result of these calculations for the one-loop correction to the effective action is zero at zero temperature. However, the direct calculation of the effective action induced by charged particles in the crossed fields that starts with the standard definition of this correction for stationary background fields as the energy of zero-point fluctuations is absent, to our knowledge. Our aim is to fill this gap and to evaluate that one-loop correction.

In performing this task, one immediately encounters with the problem that the system should be placed in a ``box'' of a finite volume in order to have a well-defined vacuum state. Otherwise, the electromagnetic potentials corresponding to such a field in the gauge where they are stationary, in particular $A_0(x)$, grow up to infinity at spatial infinity. So the work performed by this field on charged particles can be arbitrary large (in particular, larger than $2mc^2$), and the particle and antiparticle states cannot be unambiguously separated (the discussion of superstrong electric fields see, e.g., in \cite{GitGavShish,Greiner,Szpak15,MigdB}). When the system is placed in a box, its invariance under the full Poincare group is broken by this box, and the standard symmetry arguments used to prove that the one-loop correction to the effective action is zero at zero temperature are not straightforwardly applicable (see, however, Conclusion). Furthermore, the method for evaluation of the one-loop correction based on the exact solution of the Heisenberg equations for a charged particle in constant electromagnetic fields \cite{KalKazIzv,ElmfSkag95,Schwing.10} does not work either. The nontrivial boundary conditions change the commutation relations between the coordinates and momenta (see Appendix \ref{Dyson}), and the quantum equations of motion cease to be exactly solvable. Thus the direct calculations are needed.

It should be mentioned that the study of thermodynamic properties of systems in the crossed electromagnetic fields is of a peculiar importance. All the electromagnetic fields appear to be crossed for ultrarelativistic particles in the comoving reference frame. Hence, the grand thermodynamic potential we shall investigate describes the thermodynamic properties of all the ultrarelativistic systems thermalized in the comoving reference frame. The local constant field approximation that we imply in this paper is the standard tool for describing strong field effects in QED (see, e.g., \cite{Ritus.2,KimCue,BaKaStrbook,PiMuHaKermp.2}).

In the series of papers \cite{KalKaz1,KalKaz2,KalKaz3,KalKaz4}, we developed a powerful method for evaluation of the high-temperature expansion of the one-loop contribution to the effective action. It allows one to find separately the vacuum terms, the temperature and density dependent contributions, and the contribution of particles and antiparticles. In particular, these formulas allow one to find the number of particle-antiparticle pairs in the system at a given temperature and density. As for particles obeying the Bose-Einstein statistics, the high-temperature expansion of the grand thermodynamics potential (the $\Omega$-potential) reads as
\begin{widetext}
\begin{equation}\label{expan_bos}
    -\Omega_{b}(\mu)\simeq\lim_{\nu\rightarrow 0}\Big[\sum_{k,n=0}^\infty\Gamma(D-2\nu-k)\zeta(D-2\nu-k-n)\frac{\zeta^{+}_{k}(\nu)(\beta\mu)^n}{n!\beta^{D-2\nu-k}}
    +\sum^{\infty}_{l=-1}\frac{(-1)^l \zeta(-l)}{\Gamma(l+1)}\sigma^l_{\nu}(\mu)\beta^l\Big].
\end{equation}
\end{widetext}
The $\simeq$ sign indicates that the expansion in ascending powers of $\beta$ is asymptotic, and the terms exponentially suppressed in temperature ($\beta\rightarrow0$) are discarded. The spacetime dimension $D=4$ and the term with $l=-1$ should be considered as a limit, i.e., $\zeta(1)/\Gamma(0)=-1$. It is seen that at some values of $k$ and $n$ the first terms may possess singularities as $\nu\rightarrow0$ (coefficients $\zeta^{+}_k(\nu)$ are always regular), which are canceled exactly by singularities coming from $\sigma^l_{\nu}(\mu)$.

The functions $\sigma^l_{\nu}(\mu)$ and the coefficients $\zeta^{+}_k(\nu)$ entering into the expansion \eqref{expan_bos} are determined by the zeta function constructed by means of Laplacian type operator $H(\omega)$:
\begin{equation}\label{zeta_+}
    \zeta_+(\nu,\omega):=\int_{C}\frac{d\tau\tau^{\nu-1}}{2\pi i} \Tr e^{-\tau H(\omega)},
\end{equation}
where the contour $C$ runs upwards a little to the left of the imaginary axis. The operator $H(\omega)$ is a Fourier image over time of the Klein-Gordon type operator and possesses a spectrum bounded from above. The ``$+$'' index of the zeta function reminds us that its values are determined only by positive eigenvalues of $H(\omega)$.

The coefficients $\zeta^{+}_{k}(\nu)$ are the coefficients of the asymptotic expansion of the zeta function for large $\omega$:
\begin{equation}\label{zeta_asympt}
    \zeta_+(\nu,\omega)=\sum_{k=0}^N\zeta^{+}_k(\nu)\omega^{d-2\nu-k}+O(\omega^{d-2\nu-N-1}),\qquad \omega\rightarrow+\infty,
\end{equation}
and the functions $\sigma^{l}_{\nu}(\mu)$ are determined in the following way
\begin{equation}\label{def_of_sigma}
    \sigma^{l}_{\nu}(\mu)=\int_{0}^{\infty} d\omega(\omega-\mu)^l \zeta_{+}(\nu,\omega).
\end{equation}
It is the functions $\sigma^l_{\nu}(\mu)$ that contain exponentially suppressed in fields corrections and which calculation is the most difficult.

The contribution of antiparticles to the thermodynamic potential is derived from \eqref{expan_bos} by changing the sign of a chemical potential and by a simultaneous replacement $\zeta^{+}_{k}(\nu)\rightarrow \zeta^{-}_{k}(\nu)$, where the coefficients $\zeta^{-}_{k}(\nu)$ are determined from the expansion
\begin{equation}
    \zeta_+(\nu,-\omega)=\sum_{k=0}^N\zeta^{-}_k(\nu)\omega^{d-2\nu-k}+O(\omega^{d-2\nu-N-1}),\qquad \omega\rightarrow+\infty.
\end{equation}
It should be noted here that for the configuration of fields and plates under study the coefficients $\zeta^{+}_{k}(\nu)$ coincide with $\zeta^{-}_{k}(\nu)$. Hereinafter both types will be denoted by $\zeta_{k}(\nu)$.

The paper is organized as follows. Section \ref{Spectr_Sec} is devoted to the formulation of the problem, the computation of the spectral density, and the derivation of the valuable relations between the parameters of the theory that will be used for calculating $\sigma^l_{\nu}$ functions, which define the one-loop contribution to the grand thermodynamic potential. In Section \ref{Coeff_zknu_Sec}, the explicit expressions for the first six coefficients $\zeta_k(\nu)$ are found. The main trick used there is that the zeta function can be represented as an integral of the function defining the spectrum of the problem. The alternative method of calculation of the coefficients is presented in Appendix \ref{Dyson}. Section \ref{Func_snul_Sec} is devoted to the calculation of $\sigma^l_{\nu}$ functions. Despite the fact that the explicit calculation of the functions at arbitrary $l$ is impossible, emerging structures allow one to perform exact computations in the case of non-negative integer $l$. The method described allows one to derive the expressions for any $l$. The explicit calculations are carried out up to $l=3$. The last section presents the explicit expressions for finite and divergent parts of the grand thermodynamic potential and the renormalized vacuum energy taking into account the non-perturbative corrections. The non-perturbative corrections turn out to be independent of the electromagnetic fields and are exponentially suppressed at large $mL$, where $L$ is the extension of the system along the field $E$.

\section{Spectrum}\label{Spectr_Sec}

Let us consider the eigenvalue problem for the Klein-Gordon operator with the constant homogeneous crossed electromagnetic field
\begin{equation}
    A_\mu=(-Ez,Ez,0,0),\qquad \mathbf{E}=(0,0,E),\quad \mathbf{H}=(0,-E,0),
\end{equation}
where $A_\mu$ is given in the Coulomb gauge and, for definiteness, $E>0$. The naive definition of a particle is applicable in this field provided $EL<2m$, where $L$ is the size of the system along the $z$ axis (see for details \cite{Greiner}). Therefore, we impose the zero Dirichlet boundary conditions on the wave function and consider the problem on the segment $z\in[-L/2,L/2]$. Separating the variables, we obtain
\begin{equation}\label{KG_spectrum}
    H(\omega)\psi(z)=\big[\omega^2+2(\omega+p_x)Ez-p_x^2-p_y^2+\partial_z^2-m^2\big]\psi(z)=\e\psi(z),\qquad\psi(-L/2)=\psi(L/2)=0,
\end{equation}
where the particle charge is included into the definition of the electromagnetic potential. This differential equation is reduced to the Airy equation and has the general solution
\begin{equation}\label{Airy_sol}
    \psi=c_1\Ai(h)+c_2\Bi(h),\qquad h:=(2E(\omega+p_x))^{-2/3}[\e+m^2+p_y^2+(p_x-Ez)^2-(\omega+Ez)^2].
\end{equation}
The spectrum $\e$ is found as the solution to the equation
\begin{equation}\label{spectrum_eqn}
    \Ai(h_+)\Bi(h_-)-\Ai(h_-)\Bi(h_+)=0,\qquad h_{\pm}:=h\Big|_{z=\mp L/2}.
\end{equation}
Obviously, the spectral density with respect to $\e$ is equal to
\begin{equation}
\begin{split}
    \rho(\e)&=(2E(\omega+p_x))^{-2/3}|\Ai'(h_+)\Bi(h_-)-\Ai(h_-)\Bi'(h_+)+\Ai(h_+)\Bi'(h_-)-\Ai'(h_-)\Bi(h_+)|\times\\
    &\times\de(\Ai(h_+)\Bi(h_-)-\Ai(h_-)\Bi(h_+))=:(2E(\omega+p_x))^{-2/3}\vf(h_+,h_-).
\end{split}
\end{equation}

Further, we shall need the inequalities following from \eqref{KG_spectrum}. Averaging \eqref{KG_spectrum} with respect to the eigenstate, we find for particles ($\omega>0$)
\begin{equation}\label{pm_pp_rels}
\begin{split}
    \omega+p_x&=p_x-E\lan z\ran+\sqrt{(p_x-E\lan z\ran)^2+p_y^2+\lan p_z^2\ran+m^2+\e},\\
    \omega-p_x&=-2E\lan z\ran-(p_x-E\lan z\ran)+\sqrt{(p_x-E\lan z\ran)^2+p_y^2+\lan p_z^2\ran+m^2+\e}.
\end{split}
\end{equation}
Then, for $\e\geq0$,
\begin{equation}\label{ineql_lcone}
    \omega+p_x>0,\qquad \omega-p_x>-EL.
\end{equation}
Using the first inequality, we deduce from \eqref{KG_spectrum} that
\begin{equation}\label{eps_ineql}
    \e<\omega^2-p_x^2-m^2+(\omega+p_x)EL.
\end{equation}
The spectral density is zero where the above inequalities are not satisfied for $\omega>0$.

As for antiparticles ($\omega<0$), formulas \eqref{Airy_sol}, \eqref{pm_pp_rels}, \eqref{ineql_lcone}, and \eqref{eps_ineql} look as
\begin{equation}\label{antipart_ineq}
\begin{split}
    h&=(2E(|\omega|-p_x))^{-2/3}[\e+m^2+p_y^2+(p_x-Ez)^2-(|\omega|-Ez)^2],\\
    |\omega|-p_x&=-p_x+E\lan z\ran+\sqrt{(p_x-E\lan z\ran)^2+p_y^2+\lan p_z^2\ran+m^2+\e}>0,\\
    |\omega|+p_x&=2E\lan z\ran+(p_x-E\lan z\ran)+\sqrt{(p_x-E\lan z\ran)^2+p_y^2+\lan p_z^2\ran+m^2+\e}>-EL,\\
    \e&<\omega^2-p_x^2-m^2+(|\omega|-p_x)EL,
\end{split}
\end{equation}
provided $\e\geq0$.

If $\omega=0$, we have
\begin{equation}
    \e<E^2\lan z\ran^2-m^2<0,
\end{equation}
for $EL<2m$. To put it differently, $H(0)$ does not possess negative eigenvalues in this case. It also follows from \eqref{KG_spectrum} that
\begin{equation}
    \e'(\omega)=2(\omega+E\lan z\ran).
\end{equation}
Therefore, if $EL<2m$, then $\sgn(\omega)\e'(\omega)>0$ for $\e(\omega)=0$. Thus we see that, for $EL<2m$, all the applicability conditions of the formula \eqref{expan_bos} are fulfilled (see for details \cite{KalKaz2}).

\section{Coefficients $\zeta_k(\nu)$}\label{Coeff_zknu_Sec}

Despite the fact that the spectral equation \eqref{spectrum_eqn} cannot be solved explicitly, it is possible to represent the spectral zeta function as an integral of a function defining the spectrum (the so-called Gelfand-Yaglom formalism, see, e.g. \citep{Kirsten})
\begin{multline}\label{zeta_over_rho}
\zeta_+(\nu,\omega)=\frac{e^{i \pi\nu} S}{\Gamma(1-\nu)}\int_{0}^{\infty}d\e\e^{-\nu}\int\frac{d p_{x} d p_{y}}{(2\pi)^2}\rho(\varepsilon; \omega, p_x, p_y)=\frac{e^{i \pi\nu} S}{4\pi^{3/2} \Gamma(3/2-\nu)}\int_{0}^{\infty}d\e\e^{1/2-\nu}\int d p_{x}\rho(\e; \omega, p_x, 0) \\
=\frac{e^{i \pi\nu} S}{4 \pi^{3/2} \Gamma(3/2-\nu)}\int d p_{x} \frac{1}{2\pi i}\int_{\gamma}d\e\e^{1/2-\nu} \partial_{\e}\ln[\Ai(h_+)\Bi(h_-)-\Ai(h_-)\Bi(h_+)],
\end{multline}
where the contour $\gamma$ runs along the imaginary axis downwards, and $p_y$ in $h_-, h_+$ is set to zero.

Let us prove that $\zeta^+_k(\nu)=\zeta^-_k(\nu)$. It is easy to show that the spectral equation for antiparticles coincides with \eqref{spectrum_eqn} with the replacement $p_x\rightarrow-p_x$. Therefore, $\rho(\varepsilon; -\omega, p_x, p_y)=\rho(\varepsilon; \omega, -p_x, p_y)$. Due to this relation and the integral representation \eqref{zeta_over_rho}, it is clear that $\zeta(\nu,\omega)=\zeta(\nu,-\omega)$. Hence the desired equality follows.

To calculate the coefficients $\zeta_k(\nu)$, it is necessary to expand the zeta function into a series for large $\omega$, which corresponds to large negative $h_{\pm}$. Using the asymptotic expansion of Airy functions \cite{Vallee}, we obtain
\begin{multline}\label{AiryAsymp}
\Ai(h_+)\Bi(h_-)-\Ai(h_-)\Bi(h_+)\approx\\
\approx\frac{\sin\frac{2}{3}((-h_+)^{3/2}-(-h_-)^{3/2})}{\pi(h_+ h_-)^{1/4}}\Big[P\Big(\frac{2}{3}(-h_+)^{3/2}\Big)P\Big(\frac{2}{3}(-h_-)^{3/2}\Big)+Q\Big(\frac{2}{3}(-h_+)^{3/2}\Big)Q\Big(\frac{2}{3}(-h_-)^{3/2}\Big)\Big]+\\
+\frac{\cos\frac{2}{3}((-h_+)^{3/2}-(-h_-)^{3/2})}{\pi(h_+ h_-)^{1/4}}\Big[P\Big(\frac{2}{3}(-h_+)^{3/2}\Big)Q\Big(\frac{2}{3}(-h_-)^{3/2}\Big)-Q\Big(\frac{2}{3}(-h_+)^{3/2}\Big)P\Big(\frac{2}{3}(-h_-)^{3/2}\Big)\Big],
\end{multline}
where the functions $P$ and $Q$ have the following form

\begin{equation}
\begin{split}
P\Big(\frac{2}{3}z^{3/2}\Big)&=\frac{1}{\sqrt{\pi}}\sum^{\infty}_{s=0}\Big(\frac{1}{9}\Big)^{2s} \frac{\Gamma(6s+1/2)}{\Gamma(4s+1)}\frac{(-1)^s}{z^{3s}}=1-\frac{385}{4608z^3}+\cdots,\\
Q\Big(\frac{2}{3}z^{3/2}\Big)&=\frac{1}{\sqrt{\pi}}\sum^{\infty}_{s=0}\Big(\frac{1}{9}\Big)^{2s+1} \frac{\Gamma(6s+7/2)}{\Gamma(4s+3)}\frac{(-1)^s}{z^{3s+3/2}}=\frac{5}{48z^{3/2}}-\frac{85085}{663552z^{9/2}}+\cdots
\end{split}
\end{equation}
Then, for the logarithm of the expression \eqref{AiryAsymp}, we have
\begin{multline}\label{LnAiryAsymp}
\ln[\Ai(h_+)\Bi(h_-)-\Ai(h_-)\Bi(h_+)]\approx\ln\frac{\sin\frac{2}{3}((-h_+)^{3/2}-(-h_-)^{3/2})}{\pi(h_+ h_-)^{1/4}}+\\
+\ln\Big[P\Big(\frac{2}{3}(-h_+)^{3/2}\Big)-\cot\frac{2}{3}((-h_+)^{3/2}-(-h_-)^{3/2}) Q\Big(\frac{2}{3}(-h_+)^{3/2}\Big)\Big]+\\
+\ln\Big[P\Big(\frac{2}{3}(-h_-)^{3/2}\Big)+\cot\frac{2}{3}((-h_+)^{3/2}-(-h_-)^{3/2}) Q\Big(\frac{2}{3}(-h_-)^{3/2}\Big)\Big].
\end{multline}
It is taken into account that the cotangent tends to $\pm i$ on the upper and lower parts of the contour. The last two logarithms in \eqref{LnAiryAsymp} are connected by the replacement $h_+\leftrightarrow h_-$.

It is easy obtain the first several terms of the expansion of the logarithm:
\begin{multline}\label{approxAiry}
\partial_{\e}\ln[\Ai(h_+)\Bi(h_-)-\Ai(h_-)\Bi(h_+)]\approx\partial_{\e}\ln\sin\frac{2}{3}((-h_+)^{3/2}-(-h_-)^{3/2})-\frac{1}{4}\partial_{\e}\ln(h_+h_-)+\\
+\frac{5}{48}\partial_{\e}\Big(((-h_-)^{-3/2}-(-h_+)^{-3/2})\cot\frac{2}{3}((-h_+)^{3/2}-(-h_-)^{3/2})\Big)-\frac{5}{64}\partial_{\e}\Big((-h_-)^{-3}+(-h_+)^{-3}\Big)+\cdots
\end{multline}
As we will see later, the terms presented suffice to calculate $\zeta_k$ up to $k=6$.
The fact that we are interested in the explicit form of the first six coefficients
is attributable to that we managed to calculate the functions $\sigma^l_{\nu}$ till $l=3$. Thus, we shall know the expansion of the $\Omega$-potential up to $\beta^3$ (see the formula \eqref{cancel}). It should be noted that in order to derive the finite and divergent, as $\beta\rightarrow0$, parts of the expansion, it is sufficient to know the coefficients till $\zeta_4$.

Consider the first contribution
\begin{equation}
\frac{1}{2\pi i}\int_{\gamma}d\e\e^{1/2-\nu} \partial_{\e}\ln \sin\frac{2}{3}((-h_+)^{3/2}-(-h_-)^{3/2})
\end{equation}

Enclose the contour on the cut of the function $\e^{1/2-\nu}$ and perform the integral over $\e$:
\begin{multline}
\frac{\sin\pi\nu}{2E(\omega+p_x)\pi}\int^{\infty}_{0} d\e \e^{1/2-\nu}\Big((\e-m^2-(p_x-EL/2)^2+\omega^2_+)^{1/2}-(\e-m^2-(p_x+EL/2)^2+\omega^2_-)^{1/2}\Big)=\\
=\frac{\sin\pi\nu}{2E(\omega+p_x)\pi}\frac{\Gamma(3/2-\nu)\Gamma(\nu-2)}{2\pi^{1/2}}\Big[(\omega^2_- -m^2-(p_x+EL/2)^2)^{2-\nu}-(\omega^2_+ -m^2-(p_x-EL/2)^2)^{2-\nu}\Big].
\end{multline}
Here we have introduced the notation $\omega_{\pm}:=\omega\pm EL/2$. Note that the contributions with $E$ and $-E$ cannot be considered separately as the integral over $\e$ is divergent at any value of $\nu$. After integrating over $p_x$, we arrive at
\begin{equation}
-\frac{e^{i\pi\nu} S \cos\pi\nu \Gamma(\nu-5/2)}{16 E \pi^{5/2}} \frac{(\omega_-^2-m^2)^{5/2-\nu}}{\omega_-}F\Big(\frac{1}{2},1;\frac{7}{2}-\nu;\frac{\omega^2_- -m^2}{\omega_-^2}\Big)+(E\rightarrow-E).
\end{equation}
The hypergeometric function should be expanded in the vicinity of unity. The easiest way to do this is by using the relation
\begin{multline}\label{hypergeom(1-z)}
F(a,b;c;z)=\frac{\Gamma(c)\Gamma(a+b-c)}{\Gamma(a)\Gamma(b)}(1-z)^{c-a-b} F(c-a,c-b;c-a-b+1;1-z)+\\
+\frac{\Gamma(c)\Gamma(c-a-b)}{\Gamma(c-a)\Gamma(c-b)}F(a,b;a+b-c+1;1-z),\quad c-a-b \notin \mathbb{Z}.
\end{multline}
It is not difficult to see that only the second term in \eqref{hypergeom(1-z)} gives the leading contribution as $\omega\rightarrow+\infty$, so
\begin{equation}\label{zeta_1}
-\frac{e^{i\pi\nu} S}{16 E \pi^{3/2}\Gamma(5/2-\nu)} \frac{(\omega_-^2-m^2)^{5/2-\nu}}{\omega_-}\frac{1}{\nu-2} F\Big(\frac{1}{2},1;\nu-1;\frac{m^2}{\omega_-^2}\Big)+(E\rightarrow-E).
\end{equation}
Expanding the derived expression into a series in $1/\omega$, we deduce
\begin{equation}
\begin{split}
\zeta_0&\approx \frac{e^{i\pi\nu}V}{8\pi^{3/2}}\frac{1}{\Gamma(5/2-\nu)},\\
\zeta_2&\approx -\frac{e^{i\pi\nu}V}{8\pi^{3/2}}\frac{6m^2+E^2L^2(\nu-1)}{6\Gamma(3/2-\nu)},\\
\zeta_4&\approx \frac{e^{i\pi\nu}V}{8\pi^{3/2}}\frac{60m^4+20E^2L^2m^2\nu+E^4L^4\nu(\nu-1)}{120\Gamma(1/2-\nu)},\\
\zeta_6&\approx -\frac{e^{i\pi\nu}V}{8\pi^{3/2}}\frac{840m^6+420E^2L^2m^4(\nu+1)+42E^4L^4m^2\nu(\nu+1)+E^6L^6\nu(\nu^2-1)}{5040\Gamma(-1/2-\nu)}.
\end{split}
\end{equation}

Consider the term
\begin{multline}\label{loghh}
\partial_{\e}\ln\frac{1}{\pi(h_+ h_-)^{1/4}}=-\frac{1}{4}(h_+^{-1}+h_-^{-1})\partial_{\e}h_+=\\
=-\frac{1}{4}\Big(\frac{1}{\e+m^2+(p_x+EL/2)^2-\omega_-^2}+\frac{1}{\e+m^2+(p_x-EL/2)^2-\omega_+^2}\Big).
\end{multline}
The contributions with $E$ and $-E$ can be treated separately. It is convenient to enclose the contour $\gamma$ to the right and calculate the integral using residues. After being integrated, the contribution to $\zeta_+(\nu,\omega)$ reads as
\begin{equation}\label{zeta_2}
-\frac{e^{i\pi\nu} S}{16\pi \Gamma(2-\nu)}\Big(\omega_-^2-m^2\Big)^{1-\nu}+(E\rightarrow-E).
\end{equation}
Then, the contribution to $\zeta_k$ from \eqref{loghh} becomes
\begin{equation}
\begin{split}
\zeta_1\approx&-\frac{e^{i\pi\nu}S}{8\pi}\frac{1}{\Gamma(2-\nu)},\\
\zeta_3\approx&\frac{e^{i\pi\nu}S}{8\pi}\frac{4m^2+E^2L^2(2\nu-1)}{4\Gamma(1-\nu)},\\
\zeta_5\approx&-\frac{e^{i\pi\nu}S}{8\pi}\frac{m^4+\frac{1}{2}E^2L^2m^2(2\nu+1)+\frac{1}{48}E^4L^4(4\nu^2-1)}{2\Gamma(-\nu)}.
\end{split}
\end{equation}
It should be pointed out that the expressions \eqref{zeta_1} and \eqref{zeta_2} coincide exactly with the known answer for the zeta function as $E\rightarrow0$:
\begin{equation}
\zeta_+(\nu,\omega)=\frac{e^{i\pi\nu} V}{8\pi^{3/2}}\frac{(\omega^2-m^2)^{3/2-\nu}}{\Gamma(5/2-\nu)}-\frac{e^{i\pi\nu} S}{8\pi}\frac{(\omega^2-m^2)^{1-\nu}}{\Gamma(2-\nu)}.
\end{equation}

In the contribution from the last line of \eqref{approxAiry} we integrate by parts
\begin{equation}
\frac{1}{2\pi i}\int_{\gamma} d\e\e^{1/2-\nu} \partial_{\e} f(h_+, h_-)=\frac{(\nu-1/2)}{2\pi i}\int_{\gamma} d\e\e^{-\nu-1/2} f(h_+, h_-).
\end{equation}
It is sufficient to consider only the integrals of the form
\begin{multline}
\int_{\gamma} d\e \e^{-\nu-1/2} (-h_+)^{\alpha}=-2i \cos(\pi\nu)(2E(\omega+p_x))^{-\frac{2}{3}\alpha}\theta(\omega_-^2-m^2-(p_x+EL/2)^2)\times\\
\times(\omega_-^2-m^2-(p_x+EL/2)^2)^{\alpha-\nu+1/2} B(1/2-\nu,\nu-1/2-\alpha),
\end{multline}
where $\alpha<\nu-1/2<0$. In our case $\alpha=\{-\frac{3}{2};-3\}$, see \eqref{approxAiry}. The contribution with $h_-$ is obtained by the change $E\rightarrow-E$ in the final answer. The integral over $p_x$ reduces to
\begin{multline}
\int^{\infty}_{-\infty} dp (p+\omega_-)^{-\frac{2}{3}\alpha}\theta(\omega_-^2-m^2-p^2)(\omega_-^2-m^2-p^2)^{\alpha-\nu+1/2}=\\
=\sum^{\infty}_{n=0} C_{-\frac{2}{3}\alpha}^{2n} \omega_-^{-\frac{2}{3}\alpha-2n} (\omega^2_- -m^2)^{n+1+\alpha-\nu} B(n+1/2, 3/2+\alpha-\nu),
\end{multline}
where
\begin{equation}
C_n^k=\frac{\Gamma(n+1)}{\Gamma(k+1)\Gamma(n-k+1)}
\end{equation}
is the binomial coefficient, and the sum over $n$ is finite as $-\frac{2}{3}\alpha$ is a positive integer number.

As a result, the contribution to $\zeta_+(\nu,\omega)$ reads
\begin{equation}\label{zeta_3}
\frac{e^{i\pi\nu} S \cos(\pi\nu)\Gamma(\nu-1/2-\alpha)}{4\pi^{5/2}\Gamma(-\alpha)}(2E)^{-\frac{2}{3}\alpha}\sum_{n=0} C_{-\frac{2}{3}\alpha}^{2n} \omega_-^{-\frac{2}{3}\alpha-2n} (\omega^2_- -m^2)^{n+1+\alpha-\nu} B(n+1/2, 3/2+\alpha-\nu).
\end{equation}
Using this formula for the third term in \eqref{approxAiry},
\begin{equation}
((-h_-)^{-3/2}-(-h_+)^{-3/2})\cot\frac{2}{3}((-h_+)^{3/2}-(-h_-)^{3/2}),
\end{equation}
we obtain
\begin{equation}
-\frac{e^{i\pi\nu}SE}{\pi^{3/2}\Gamma(1/2-\nu)}\omega_-(\omega_-^2-m^2)^{-\nu-1/2}+(E\rightarrow-E).
\end{equation}
Expanding this expression into a series in $1/\omega$, we arrive at
\begin{equation}
\begin{split}
\zeta_4&\approx -\frac{e^{i\pi\nu}V}{8\pi^{3/2}} \frac{16E^2\nu}{\Gamma(1/2-\nu)},\\
\zeta_6&\approx \frac{e^{i\pi\nu}V}{8\pi^{3/2}} \frac{8(\nu+1)(6m^2E^2+E^4L^2\nu)}{3\Gamma(-1/2-\nu)}.
\end{split}
\end{equation}

For the fourth term in \eqref{approxAiry},
\begin{equation}
(-h_-)^{-3}+(-h_+)^{-3},
\end{equation}
we obtain the following contribution to the zeta function
\begin{equation}
-\frac{e^{i\pi\nu} S E^2}{4\pi \Gamma(-\nu)}\Big[2(\nu+1)\omega_-^2(\omega_-^2-m^2)^{-\nu-2}-(\omega_-^2-m^2)^{-\nu-1}\Big]+(E\rightarrow-E).
\end{equation}
The contribution to $\zeta_5$
\begin{equation}
\zeta_5\approx -\frac{e^{i\pi\nu}SE^2}{2\pi} \frac{2\nu+1}{\Gamma(-\nu)}.
\end{equation}

Collecting all the contributions together and taking the coefficients into account, we find
\begin{equation}\label{zeta_k}
\begin{split}
	\zeta_0(\nu)&= \frac{e^{i\pi\nu}V}{8\pi^{3/2}}\frac{1}{\Gamma(5/2-\nu)},\\
	\zeta_1(\nu)&= -\frac{e^{i\pi\nu}S}{8\pi}\frac{1}{\Gamma(2-\nu)},\\
	\zeta_2(\nu)&= -\frac{e^{i\pi\nu}V}{8\pi^{3/2}}\frac{1}{\Gamma(3/2-\nu)}\Big[m^2+\frac{1}{6}(\nu-1)(EL)^2\Big],\\
	\zeta_3(\nu)&= \frac{e^{i\pi\nu}S}{8\pi}\frac{1}{\Gamma(1-\nu)}\Big[m^2+\frac{1}{2}(\nu-\frac{1}{2})(EL)^2\Big],\\
	\zeta_4(\nu)&= \frac{e^{i\pi\nu}V}{8\pi^{3/2}}\frac{1}{\Gamma(1/2-\nu)}\Big[\frac{1}{2}m^4+\frac{1}{6}\nu(EL)^2m^2-\frac{5}{3}\nu E^2+\frac{1}{120}\nu(\nu-1)(EL)^4\Big],\\
	\zeta_5(\nu)&= -\frac{e^{i\pi\nu}S}{8\pi}\frac{1}{\Gamma(-\nu)}\Big[\frac{1}{2}m^4+\frac{1}{2}(\nu+\frac{1}{2})(EL)^2m^2-\frac{5}{8}(\nu+\frac{1}{2})E^2+\frac{1}{24}(\nu+\frac{1}{2})(\nu-\frac{1}{2})(EL)^4\Big],\\
	\zeta_6(\nu)&= -\frac{e^{i\pi\nu}V}{8\pi^{3/2}}\frac{1}{\Gamma(-1/2-\nu)}\Big[\frac{1}{6}m^6+\frac{1}{12}(\nu+1)(EL)^2m^4-\frac{5}{3}(\nu+1)m^2E^2+\frac{1}{120}\nu(\nu+1)(EL)^4m^2-\\
	&-\frac{5}{18}\nu(\nu+1)E^4L^2+\frac{1}{5040}\nu(\nu-1)(\nu+1)(EL)^6\Big].
\end{split}
\end{equation}
The above mentioned expressions for $\zeta_k(\nu)$ allow one to conjecture the general structure at arbitrary $k$:
\begin{equation}\label{zeta_k_general}
\zeta_k(\nu)=\frac{1}{\Gamma(5/2-k/2-\nu)}\sum^{[k/2]}_{n=0}\alpha_n \nu^n.
\end{equation}
The rigorous proof of \eqref{zeta_k_general} follows from the analysis of the expressions \eqref{zeta_1},\eqref{zeta_2}, and \eqref{zeta_3}. An alternative method of calculation of the coefficients by the use of Dyson series is represented in Appendix~\ref{Dyson}.

\section{Functions $\s_\nu^{l}(\mu)$}\label{Func_snul_Sec}

In accordance with the general formulas (see Introduction), for the high-temperature expansion to be obtained, one needs to find the expression for the function
\begin{equation}\label{sigma_Ai_ini}
    \s_\nu^l(\mu):=\frac{e^{i\pi\nu}S}{\Ga(1-\nu)}\int_0^\infty d\omega(\omega-\mu)^l\int_0^\infty d\e\e^{-\nu}\int\frac{dp_xdp_y}{(2\pi)^2}\rho(\e;\omega,p_x,p_y),
\end{equation}
where $S:=L_xL_y$, in the form of an analytic function of $\nu$ and $l$. It was shown in \cite{KalKaz1} that the integral over $\e$ converges when
\begin{equation}\label{converg_dom1}
    \re\nu<1.
\end{equation}
The integral over $\omega$ converges for
\begin{equation}\label{converg_dom2}
    \re(\nu-l/2)>(d+1)/2=2.
\end{equation}
Therefore, it is useful to calculate \eqref{sigma_Ai_ini} in the region \eqref{converg_dom1}, \eqref{converg_dom2}, where the multiple integral \eqref{sigma_Ai_ini} converges, and then to continue $\s^l_\nu(\mu)$ by analyticity to the required ``physical'' values of the parameters $\nu$ and $l$. Recall that, according to the Hartogs theorem (see, e.g., \cite{Shabat}), the function that is analytic with respect to each variable is analytic with respect to all of them. The uniqueness of analytical continuation also holds for such functions. As follows from the general analysis given in \cite{KalKaz1}, the function $\s^l_\nu(\mu)$ is a meromorphic function of $\nu$ and $l$. It possesses the singularities in the form of simple poles at
\begin{equation}\label{sigma_poles}
    d+l-2\nu+2\in\N,\qquad d=3,
\end{equation}
provided that there is a neighborhood of the point $\omega=0$ that does not contain the points of the particle's energy spectrum and $\mu$ belongs to this neighborhood.

First, we integrate \eqref{sigma_Ai_ini} over $p_y$. After the replacement $\e\rightarrow\e-p_y^2$, the spectral density $\rho(\e)$ becomes independent of $p_y$, and the integral over $p_y$ is reduced to
\begin{equation}
    \int dp_y\frac{\theta(\e-p_y^2)}{(\e-p_y^2)^\nu}=\theta(\e)\e^{1/2-\nu}\frac{\sqrt{\pi}\Ga(1-\nu)}{\Ga(3/2-\nu)},\qquad\re\nu<1.
\end{equation}
Further, we make the integration variables dimensionless
\begin{equation}\label{redefn1}
    \omega\rightarrow m\omega,\qquad p_x\rightarrow mp_x,\qquad\e\rightarrow m^2\e,
\end{equation}
introduce convenient notation
\begin{equation}\label{conven_notion}
    \bar{m}^2:=m^2/E,\qquad w:=EL/(2m),\qquad c:=\bar{m}^2w=mL/2\qquad\bar{\mu}:=\mu/m,
\end{equation}
and pass to the light-cone variables
\begin{equation}\label{redefn2}
    u:=\omega+p_x,\qquad v:=\omega-p_x,\qquad d\omega dp_x=\frac12 dudv.
\end{equation}
The region of integration with respect to these variables is determined by the inequalities \eqref{ineql_lcone}. Stretching the variables
\begin{equation}\label{redefn3}
    v\rightarrow 2wv,\qquad u\rightarrow\bar{m}^2u/2,\qquad\e\rightarrow c\e,
\end{equation}
we arrive at
\begin{equation}\label{sigma_l_nu1}
\begin{split}
    \s^l_\nu(\mu)&=\frac{e^{i\pi\nu}Sm^{3+l-2\nu}}{8\pi^{3/2}2^l\Ga(3/2-\nu)}c^{5/2-\nu}\int_0^\infty duu^{-2/3}\int_{-1}^\infty dv\Big(\frac{\bar{m}^2u}{2}+2wv-2\bar{\mu}\Big)^l\times\\
    &\times\int_0^\infty d\e\e^{1/2-\nu}\vf\big(cu^{-2/3}(\e+c^{-1}-uv+u),cu^{-2/3}(\e+c^{-1}-uv-u)\big).
\end{split}
\end{equation}
As seen from this expression, we can integrate over the variable $v$ as it was done above with the variable $p_y$.

To this aim, we shift the integration variable
\begin{equation}\label{redefn4}
    \e\rightarrow\e-c^{-1}+uv.
\end{equation}
Then the integrand of \eqref{sigma_l_nu1} includes
\begin{equation}\label{prod_thet}
    \theta(v+1)\theta\big(v+(\e-c^{-1})/u\big).
\end{equation}
On making the redefinitions \eqref{redefn1}, \eqref{redefn2}, \eqref{redefn3}, and \eqref{redefn4}, the inequality \eqref{eps_ineql} has the form
\begin{equation}\label{eps_ineql2}
    \e<u.
\end{equation}
Therefore, the first $\theta$-function in \eqref{prod_thet} can be removed. As a result, the integral over $v$ becomes
\begin{multline}
    \int_{-\infty}^\infty dv\theta(\e+uv-c^{-1})(\e+uv-c^{-1})^{1/2-\nu}\Big(\frac{\bar{m}^2u}{2}+2wv-2\bar{\mu}\Big)^l=\frac{\Ga(3/2-\nu)\Ga(\nu-l-3/2)}{u\Ga(-l)}\times\\
    \times \Big(\frac{2w}{u}\Big)^l\big(c^{-1}-\e-\bar{\mu}u/w+\bar{m}^2u^2/(4w)\big)^{3/2+l-\nu},\qquad \re\nu<3/2,\;\;\re(\nu-l)>3/2,
\end{multline}
where it is assumed that $(\bar{\mu}-w)^2<1$. Stretching the integration variable
\begin{equation}
    \e\rightarrow u\e,
\end{equation}
and taking into account the inequality \eqref{eps_ineql2}, we obtain
\begin{multline}\label{sigma_l_nu2}
    \s^l_\nu(\mu)=c^{5/2-\nu}\Ga(\nu-l-3/2)\frac{e^{i\pi\nu}Sm^{3+l-2\nu}}{8\pi^{3/2}w^{-l}\Ga(-l)}\times\\
    \times\int_0^\infty \frac{du}{u^{l+2/3}}\int_{-\infty}^1 d\e\Big(c^{-1}-\e u-\frac{\bar{\mu}u}{w}+\frac{cu^2}{4w^2}\Big)^{3/2+l-\nu}\vf\big(cu^{1/3}(\e+1),cu^{1/3}(\e-1)\big).
\end{multline}
Further simplification of this integral is impossible without knowledge of the explicit expression for the function $\vf$. It appears at first sight that \eqref{sigma_l_nu2} possesses singularities at
\begin{equation}
    l-\nu+5/2\in\N,
\end{equation}
which contradicts the general statement \eqref{sigma_poles}. However, for such values of $\nu$ and $l$, the integral on the second line of \eqref{sigma_l_nu2} understood in the sense of analytic continuation goes to zero (see \eqref{integral2}) and, consequently, $\s^l_\nu(\mu)$ is regular at these points.

\subsection{Functions $\s_\nu^{l}(\mu)$ for nonnegative integer $l$}

The presence of $\Ga(-l)$ in the denominator of \eqref{sigma_l_nu2} allows one to obtain the exact expression for $\s^l_\nu(\mu)$ at $l=\overline{0,\infty}$. To this end, it is sufficient to investigate the singularities of the integral on the second line in \eqref{sigma_l_nu2} in the complex $l$ plane near $l=\overline{0,\infty}$. As the theorem \ref{anal_int_prop} shows, these singularities are the poles with the residues found by expansion of the integrand in the asymptotic series. There is no need to evaluate the integral. Namely, (see for details \cite{GSh,ParKam01,Wong})
\begin{thm}\label{anal_int_prop}
Let $\vf(x)$ be absolutely integrable on $(0,\La]$ and the following asymptotic expansion takes place
\begin{equation}\label{asympt_exp}
    \vf(x)=\sum_{k=0}^N a_k x^k+O(x^{N+1}),
\end{equation}
for $x\rightarrow+0$. Then the function
\begin{equation}\label{Ila}
    I(\la)=\int_0^\La dx x^\la\vf(x),\quad 0<\La<+\infty,
\end{equation}
is analytic for $\re\la>-1$ and can be analytically continued to the region $\re\la>-2-N$, where it possesses the simple poles at the points $\la=-k$, $k=\overline{1,N+1}$, with the residues $a_{k-1}$, respectively.
\end{thm}

In order to apply the theorem to \eqref{sigma_l_nu2}, we pass from the integration variable $\e$ to
\begin{equation}
    \xi^{-1}:=1/(cu)-\e-\bar{\mu}/w+cu/(4w^2).
\end{equation}
It is not difficult to see that
\begin{equation}
    \xi\in[0,\xi_0],\qquad\xi_0:=w/(1-w-\bar{\mu})>0,
\end{equation}
where it is supposed that $\mu+w<1$. Let us stretch the integration variable
\begin{equation}
    u\rightarrow\xi u.
\end{equation}
Then the integral on the second line of \eqref{sigma_l_nu2} can be cast into the form
\begin{equation}\label{g_function}
    \int_0^{\xi_0}d\xi\xi^{-l-1}g(\xi),\qquad g(\xi):=\xi^{-2/3}\int_{u_-}^{u_+}duu^{5/6-\nu}\vf(h_+,h_-),
\end{equation}
where
\begin{equation}\label{h_pm}
    h_\pm=cu^{1/3}\xi^{1/3}(\e\pm1)=cu^{1/3}\xi^{-2/3}\Big(\frac{1}{cu}-1-\frac{\bar{\mu}\xi}{w}+\frac{cu}{4w^2}\xi^2\pm\xi\Big),
\end{equation}
and
\begin{equation}
    cu_\pm=\frac{2w^2}{\xi^2}\Big(1+\big(1+\frac{\bar{\mu}}{w}\big)\xi\Big)\bigg\{1\pm\bigg[1-\frac{\xi^2}{w^2\big(1+\big(1+\frac{\bar{\mu}}{w}\big)\xi\big)^2}\bigg]^{1/2}\bigg\}.
\end{equation}
The last expression is obtained from the solution of the equation $h_-=0$. The function $g(\xi)$ is bounded on the integration interval except possibly the vicinity of the point $\xi=0$. Therefore, in order to evaluate  \eqref{sigma_l_nu2} at $l=\overline{0,\infty}$, it is sufficient to derive the asymptotic expansion of the integrand for $\xi\rightarrow+0$.

%At that the constraint \eqref{converg_dom1} can be relaxed можно ослабить, считая, что переменная $\nu$ находится в некоторой области аналитичности $\s^l_\nu(\mu)$, а затем продолжить результат по аналитичности на всю комплексную $\nu$-плоскость (конечно, за исключением точек, где  $\s^l_\nu(\mu)$ имеет полюс).

For $\xi\rightarrow0$, it is useful to split the integration region of the variable $u$ in \eqref{g_function} into the three intervals:
\begin{equation}\label{ABC_intervals}
    [u_-,u_+]=[u_-,\bar{u}_-]\cup[\bar{u}_-,\bar{u}_+]\cup[\bar{u}_+,u_+]=:A\cup B\cup C,
\end{equation}
where
\begin{equation}\label{u_pm_bar}
    c\bar{u}_\pm=\frac{2w^2}{\xi^2}\Big(1+\big(\e_0+\frac{\bar{\mu}}{w}\big)\xi\Big)\bigg\{1\pm\bigg[1-\frac{\xi^2}{w^2\big(1+\big(\e_0+\frac{\bar{\mu}}{w}\big)\xi\big)^2}\bigg]^{1/2}\bigg\}, \qquad\e_0<-1.
\end{equation}
The boundaries of the integration region \eqref{u_pm_bar} are obtained from the solution of the equation $\e=\e_0$, where $\e_0$ is some constant independent of $\xi$. For the intervals $A$ and $C$, the quantity $\e\in[\e_0,1]$. For the interval $B$, both $h_+$ and $h_-$ are nonnegative as long as $\e<\e_0$.

\paragraph{Intervals $A$ and $C$.}

On the interval $A$, the integration variable $u\rightarrow c^{-1}$ for $\xi\rightarrow0$. In this case, Eq. \eqref{spectrum_eqn} has the form
\begin{equation}\label{spectrum_eqn_A}
    \Ai(h_+)\Bi(h_-)-\Ai(h_-)\Bi(h_+)\approx-\frac{2}{\pi}(cu^{1/3}\xi^{1/3}+2c^4u^{4/3}\e\xi^{4/3})=0,
\end{equation}
where $\e\in[\e_0,1]$. Then it follows from \eqref{spectrum_eqn_A} that
\begin{equation}
    u=0,\quad\text{or}\quad u\sim1/\xi.
\end{equation}
However, $u\approx c^{-1}$. Consequently, for $\xi\rightarrow0$, Eq. \eqref{spectrum_eqn} does not possess solutions on the interval $A$.

On the interval $C$, the integration variable $u\sim\xi^{-2}$ for $\xi\rightarrow0$. Hence, it is useful to redefine the integration variable $u\rightarrow\xi^{-2}u$. Then the integration limits become
\begin{equation}%\label{interval_C}
    \xi^2\bar{u}_+\approx \frac{4w^2}{c} \Big[1+\Big(\e_0+\frac{\bar{\mu}}{w}\Big)\xi\Big], \qquad \xi^2u_+ \approx \frac{4w^2}{c} \Big[1+\Big(1+\frac{\bar{\mu}}{w}\Big)\xi\Big],
\end{equation}
for $\xi\rightarrow0$. On stretching $u$, the additional factor $\xi^{2\nu}$ appears in the expression \eqref{g_function} for the function $g(\xi)$ (on the interval $C$). If Eq. \eqref{spectrum_eqn} has solutions on the interval $C$, then the integration over $u$ in \eqref{g_function} is removed and $u=4w^2/c+o(1)$, where $o(1)$ does not contain the powers of $\xi^{\nu}$. Then, in developing $g(\xi)$ as a series in $\xi$, the factor $\xi^{2\nu}$ cannot be canceled out, i.e., the expansion of $g(\xi)$ (on the interval $C$) in the vicinity of $\xi=0$ does not contain integer powers of $\xi$. Consequently, as follows from the theorem \ref{anal_int_prop} and the form of the integral over $\xi$ in \eqref{g_function}, the interval $C$ does contribute to the poles at nonnegative integer powers of $l$.

%Отметим, что границы интервалов $A$ и $C$ можно расширить так, чтобы на границе $\e=\e_0<-1$, где $\e_0$ не зависит от $\xi$. Значение \eqref{u_pm_bar} изменяется при этом очевидным образом. Анализ вкладов от интервалов $A$ и $C$, изложенный выше, остается в силе и для измененных $A$ и $C$.

\paragraph{Interval $B$.}

On the interval $B$, the arguments of the Airy functions entering into the equation specifying the spectrum \eqref{spectrum_eqn} are negative and tend to $-\infty$ for $\xi\rightarrow+0$. Therefore, we can employ the asymptotic expansion of the Airy functions for large negative arguments and solve Eq. \eqref{spectrum_eqn} with respect to $u$, bearing in mind that $\xi\rightarrow0$. The first three terms of the expansion with respect to $\xi$ are written as
\begin{equation}\label{y_n}
    cu_n=\Big(1+\frac{\pi^2n^2}{4c^2}\Big) \bigg\{1-\frac{\bar{\mu}}{w}\xi+\Big[\frac{\bar{\mu}^2}{w^2}+\Big(1+\frac{\pi^2n^2}{4c^2}\Big)\frac{3\pi^4n^4+4w^2c^2(\pi^2n^2-15)}{12\pi^4w^2n^4}\Big]\xi^2+O(\xi^3)\bigg\}, \quad n=\overline{1,N(\xi)},
\end{equation}
where $N(\xi)\rightarrow\infty$ for $\xi\rightarrow+0$. For example, in the leading order, we obtain from \eqref{spectrum_eqn} that
\begin{equation}
    \Big(1+\frac{\bar{\mu}\xi}{w}-\frac{1}{cu_n}-\frac{cu_n}{4w^2}\xi^2-\xi\Big)^{3/2} -\Big(1+\frac{\bar{\mu}\xi}{w}-\frac{1}{cu_n}-\frac{cu_n}{4w^2}\xi^2+\xi\Big)^{3/2} \approx\frac{3\pi n}{2c^{3/2}}\xi u_n^{-1/2}.
\end{equation}
The solution of this equation reproduces the leading term of the expansion \eqref{y_n}. The integral over $u$ in \eqref{g_function} is reduced to the sum over the roots of Eq. \eqref{spectrum_eqn} since on the interval $C$ and for $\xi\rightarrow+0$:
\begin{equation}\label{vf_PM}
    \vf(h_+,h_-)=\sum_{n=1}^\infty\Big|\frac{\Ai'(h_+)\Bi(h_-)-\Ai(h_-)\Bi'(h_+)+\Ai(h_+)\Bi'(h_-)-\Ai'(h_-)\Bi(h_+)}{h'_+[\Ai'(h_+)\Bi(h_-)-\Ai(h_-)\Bi'(h_+)]+h'_-[\Ai(h_+)\Bi'(h_-)-\Ai'(h_-)\Bi(h_+)]}\Big|\de(u-u_n(\xi)),
\end{equation}
where $h'_\pm=\partial_uh_\pm$. Substituting the expression \eqref{y_n} for $u_n$ into the resulting sum and expanding the outcome in a series with respect to $\xi$, we have
\begin{equation}
    g(\xi)=c^{\nu-5/2}\sum_{n=1}^\infty\Big(1+\frac{\pi^2 n^2}{4c^2}\Big)^{3/2-\nu}[b_0+b_1\xi+b_2\xi^2+O(\xi^3)],
\end{equation}
where
\begin{equation}
\begin{split}
    b_0&=1,\qquad b_1=(\nu-5/2)\frac{\bar{\mu}}{w},\\
    b_2&=(\nu-7/2)\Big\{(\nu-5/2)\frac{\bar{\mu}^2}{2w^2}-\Big(1+\frac{\pi^2 n^2}{4c^2}\Big)\Big[\frac{1}{4w^2}-\frac{c^2(4c^2+15)}{3\pi^4n^4} +\frac{4c^4}{3\pi^4n^4}\Big(1+\frac{\pi^2 n^2}{4c^2}\Big) \Big]\Big\}.
\end{split}
\end{equation}
Thus, employing the theorem \ref{anal_int_prop}, we deduce
\begin{equation}
    \frac{(-1)^l}{\Ga(l+1)}\s^l_\nu(\mu)=\Ga(\nu-l-3/2)\frac{e^{i\pi\nu}S}{8\pi^{3/2}}m^{3+l-2\nu}w^l\sum_{n=1}^\infty\Big(1+\frac{\pi^2 n^2}{4c^2}\Big)^{3/2-\nu} b_l,\qquad l=\overline{0,\infty}.
\end{equation}
The explicit expressions for the sums over $n$ are presented in Appendix \ref{Some_Series}. Using these expressions, we can write
\begin{equation}\label{sigma_expl}
\begin{split}
    \s^0_\nu(\mu)=\, & e^{i\pi\nu}\frac{SLm^{4-2\nu}}{16\pi^2}\Big[\Ga(\nu-2)-\frac{\sqrt{\pi}}{2c}\Ga(\nu-3/2)+\frac{2\sqrt{\pi}}{\Ga(5/2-\nu)}T^0_{3/2-\nu}(4c)\Big],\\
    \s^1_\nu(\mu)=\, & -\mu \s^0_\nu(\mu)=-\mu \s^0_\nu(0),\\
%   \s^2_\nu(\mu)=\, & \mu^2 \s^0_\nu(0)-e^{i\pi\nu}\frac{SLm^{6-2\nu}}{32\pi^2}%\Big[\Ga(\nu-3)-\frac{\sqrt{\pi}}{2c}\Ga(\nu-5/2)+\frac{2\sqrt{\pi}}{\Ga(7/2-\nu)}%T^0_{5/2-\nu}(4c)-\\
%   &-\frac{2w^4}{3}\Ga(\nu-2)+\frac{5\sqrt{\pi}w^2}{6c}\Ga(\nu-3/2)-\frac{5w^4}{3c^2}%\Ga(\nu-1)-\\
%    &-\sqrt{\pi}\frac{w^2}{3c^2}\frac{(4c^2+15)T^4_{5/2-\nu}(4c)+4c^2T^4_{7/2-\nu}%(4c) }{\Ga(7/2-\nu)} \Big].\\
	\s^2_\nu(\mu)=\, & \mu^2 \s^0_\nu(0)-e^{i\pi\nu}\frac{SLm^{6-2\nu}}{32\pi^2}\Big[\Ga(\nu-3)-\frac{\sqrt{\pi}}{2c}\Ga(\nu-5/2)+\frac{2\sqrt{\pi}}{\Ga(7/2-\nu)}T^0_{5/2-\nu}(4c)-\\
	 &-\frac{2w^2}{3}\Ga(\nu-2)+\frac{\sqrt{\pi}w^2}{c}\Ga(\nu-3/2)-\frac{5w^2}{3c^2}\Ga(\nu-1)+\frac{5w^2\sqrt{\pi}}{16c^3}\Ga(\nu-1/2)-\\
    &-\sqrt{\pi}\frac{w^2}{3c^2}\frac{(4c^2+15)T^4_{5/2-\nu}(4c)+4c^2T^4_{7/2-\nu}(4c) }{\Ga(7/2-\nu)} \Big].
\end{split}
\end{equation}
The functions $T^k_\al(4c)$ are exponentially suppressed for large $c$.

We see from formulas \eqref{sigma_Ai_ini}, \eqref{sigma_expl} that $\s^1_\nu(0)$ vanishes. In general, for the system at issue
\begin{equation}\label{sigma_prop}
    \s^{2k+1}_\nu(0)=0,\qquad k=\overline{0,\infty}.
\end{equation}
Indeed, it follows from \eqref{h_pm} that
\begin{equation}\label{xi_minus_xi}
    h_\pm\underset{\xi\rightarrow-\xi}{\longrightarrow}h_\mp,\qquad h'_\pm\underset{\xi\rightarrow-\xi}{\longrightarrow}h'_\mp,
\end{equation}
for $\mu=0$. As long as Eq. \eqref{spectrum_eqn} remains unchanged under \eqref{xi_minus_xi} and $\vf(h_+,h_-)=\vf(h_-,h_+)$ (see \eqref{vf_PM}), the expansion of $g(\xi)$ with respect to $\xi$ contains only even powers of $\xi$. Therefore, we have \eqref{sigma_prop}.

It is clear from \eqref{def_of_sigma} that
\begin{equation}
    \sigma^l_{\nu}(\mu)=\sum_{n=0}^{l}C^n_l \sigma^n_{\nu}(0)(-\mu)^{l-n},\qquad C_n^k=\frac{n!}{k!(n-k)!}.
\end{equation}
Then formula \eqref{sigma_prop} implies that
\begin{equation}
    \s^{2k+1}_\nu(\mu)=-\s^{2k+1}_\nu(-\mu).
\end{equation}
Also we obtain
\begin{equation}\label{sigma_3}
    \s^3_\nu(\mu)=-3\mu\s^2_\nu(0)-\mu^3\s^0_\nu(0).
\end{equation}
Taking into account the inequalities \eqref{antipart_ineq}, it is not difficult to check that the contribution of antiparticles to $\s^l_\nu(\mu)$ is equal to the contribution of particles with $\mu\rightarrow-\mu$. Hence, the high-temperature expansion of the thermodynamic potential including the contributions of particles and antiparticles does not contain $\s^{2k+1}_\nu(\mu)$. The terms with $\s^{2k}_\nu(\mu)$, $k=\overline{1,\infty}$, do not contribute to the high-temperature expansion either (see \eqref{expan_bos}). Consequently, the one-loop contribution to the $\Omega$-potential coming from particles and antiparticles does not include the second term with  $l=\overline{1,\infty}$. Notice that, in contrast to the system of charged particles in a homogeneous magnetic field (see \cite{KalKaz3}), the functions $\s^l_\nu(\mu)$, $l=\overline{0,\infty}$, do not contain the terms that are nonanalytic with respect to the coupling constant or the external field. Such contributions do not arise in perturbative in $\xi$ solution of Eq. \eqref{spectrum_eqn} with respect to $u$. As for $\s^l_\nu(\mu)$, $l=\overline{0,3}$, this property is seen directly from the expressions \eqref{sigma_expl}, \eqref{sigma_3}.

\subsection{Function $\s_\nu^{-1}(\mu)$}

In order to find the high-temperature expansion for the thermodynamic potential of bosons, we need to derive $\s_\nu^{-1}(\mu)$. The considerations of the previous subsection are not applicable in this case. We did not succeed in finding the exact expression for $\s_\nu^{-1}(\mu)$, but we managed to find it under the assumption that (see \eqref{conven_notion})
\begin{equation}\label{approx_cond}
    |w|<1,\qquad c\gg1.
\end{equation}
Let us change the integration variable in the integral on the second line of \eqref{sigma_l_nu2},
\begin{equation}
    u\rightarrow c^{-1}u,
\end{equation}
and rewrite it as a contour integral
\begin{equation}\label{integral1}
    c^{\nu-5/2}\int_0^\infty\frac{du}{u^{l+1}}\int_C\frac{d\e}{2\pi i}\Big(1-\e u-\frac{\bar{\mu}}{w}u+\frac{u^2}{4w^2}\Big)^{3/2+l-\nu}\partial_\e\ln(\Ai(h_+)\Bi(h_-)-\Ai(h_-)\Bi(h_+)),
\end{equation}
where the contour $C$ goes from $-\infty$ a little bit lower than the real axis, encircles the origin, and then runs to $-\infty$ a little bit higher than the real axis, $h_\pm=c^{2/3}u^{1/3}(\e\pm1)$, and the principal branches of the multivalued functions are taken.

The logarithmic derivative entering into \eqref{integral1} does not possess singularities out of the negative part of the real axis and tends to zero for $|\e|\rightarrow\infty$. Therefore, taking $\re\nu$ sufficiently large, we can deform the contour $C$ and reduce the integral \eqref{integral1} to the integral over the branch cut of the power function in \eqref{integral1}. As a result, we obtain
\begin{equation}\label{integral2}
    c^{\nu-5/2}\frac{\sin\pi(\nu-l-3/2)}{\pi} \int_0^\infty\frac{du}{u^{\nu-1/2}} \int^\infty_0 d\e\e^{3/2+l-\nu}\partial_\e\ln(\Ai(h_+)\Bi(h_-)-\Ai(h_-)\Bi(h_+)),
\end{equation}
where
\begin{equation}
    h_\pm=c^{2/3}u^{1/3}\Big(\e+\frac{1}{u}-\frac{\bar{\mu}}{w}+\frac{u}{4w^2}\pm1\Big).
\end{equation}
Bearing in mind the conditions \eqref{approx_cond}, it is not difficult to see from the asymptotic behavior of the Airy functions that, in the given integration region,
\begin{equation}
    \frac{\Ai(h_+)\Bi(h_-)}{\Ai(h_-)\Bi(h_+)}
\end{equation}
is exponentially small. Therefore, up to exponentially suppressed terms at $(c/w)\rightarrow\infty$, we have
\begin{equation}\label{int_approx}
    \partial_\e\ln(\Ai(h_+)\Bi(h_-)-\Ai(h_-)\Bi(h_+))\approx\partial_\e\ln\Ai(h_-)+\partial_\e\ln\Bi(h_+).
\end{equation}
Unfortunately, even in this case, we did not succeed in exact evaluation of the integral \eqref{integral2}.

Below, we shall derive the leading in $c/w$ contribution to $\s^{-1}_\nu(\mu)$, the terms diverging for $\nu\rightarrow0$, and take into account the asymptotics of $\s^{-1}_\nu(\mu)$ in the vicinity of the singular point in the complex $\mu$ plane where the chemical potential approaches to the lowest energy level. The integral \eqref{integral2} possesses the branch point in the $\mu$ plane when the pole of the logarithmic derivative tends to the point $\e=0$.

We start with the leading contribution and the terms singular at $\nu\rightarrow0$. Developing the logarithmic derivative as an asymptotic series, we deduce
\begin{equation}\label{asympt_log}
    \partial_\e\ln(\Ai(h_+)\Bi(h_-)-\Ai(h_-)\Bi(h_+))\approx cu^{1/2}(\tilde{h}^{1/2}_+-\tilde{h}^{1/2}_-) -\frac14[\tilde{h}^{-1}_+ +\tilde{h}^{-1}_-]+\cdots,
\end{equation}
where $\tilde{h}_\pm:=c^{-2/3}u^{-1/3}h_\pm$. This asymptotic expansion has the same form for both the exact logarithmic derivative and the approximate expression \eqref{int_approx}. The terms presented in \eqref{asympt_log} are sufficient to find all the contributions diverging at $\nu\rightarrow0$ for $l=-1$. On substituting the expansion \eqref{asympt_log} into \eqref{integral2}, the integrals over $\e$ of every term in the series are reduced to the beta function:
\begin{equation}
    \int^\infty_0 d\e\e^{3/2+l-\nu}h_{\pm}^\alpha=\Big(\frac{1}{u}-\frac{\bar{\mu}}{w}+\frac{u}{4w^2}\pm1\Big)^{5/2+l-\nu+\alpha}\frac{\Ga(l-\nu+5/2)\Ga(\nu-l-5/2-\al)}{\Ga(-\al)}.
\end{equation}
Having stretched the integration variable $u\rightarrow2wu$, the integral over $u$ becomes (see \cite{PrBr1})
\begin{equation}\label{hypergeom_int}
\begin{split}
    \int_0^\infty\frac{dx x^{\be-1}}{(x^2+2bx+1)^\rho}=\,&\frac{\Ga(\be/2)\Ga(\rho-\be/2)}{2\Ga(\rho)}F(\be/2,\rho-\be/2;1/2;b^2)-\\ &-\frac{\Ga((\be+1)/2)\Ga(\rho-\be/2+1/2)}{\Ga(\rho)}bF((\be+1)/2,\rho-\be/2+1/2;3/2;b^2).
\end{split}
\end{equation}
Then the leading term of the expansion \eqref{asympt_log} gives the contribution to $\s^{-1}_\nu(\mu)$ of the form
\begin{equation}
    e^{i\pi\nu}\frac{m^{3-2\nu}}{12\pi}S\int_{-L/2}^{L/2}dz\Big\{(1-\tilde{\mu}^2)^{3/2}+\frac{3\tilde{\mu}}{2\pi}\Ga(\nu-1)+\frac{\tilde{\mu}^3}{\pi}\Ga(\nu)+\frac2{\pi} \Big[(1-\tilde{\mu})^{3/2}\arcsin\tilde{\mu}+\tilde{\mu}\Big(\frac43\tilde{\mu}^2-1\Big) \Big] \Big\},
\end{equation}
where $\tilde{\mu}:=\bar{\mu}+Ez/m$ and, for convenience, the expression is written as the integral over $z$. As for the next term of the asymptotic expansion \eqref{asympt_log}, we will obtain only the contribution that is singular at $\nu\rightarrow0$. Using the above integrals, it is easy to see that this contribution to $\s^{-1}_{\nu}(\mu)$ is the pole part of
\begin{equation}
    -e^{i\pi\nu}\frac{m^{2-2\nu}}{32\pi}S\Ga(\nu-1)\{2+(\nu-1)[(\bar{\mu}+w)^2+(\bar{\mu}-w)^2]\}=-\frac{Sm^2}{16\pi}\frac{\bar{\mu}^2+w^2-1}{\nu}+\cdots.
\end{equation}

Thus, it only remains for us to find the asymptotics of $\s^{-1}_\nu(\mu)$ in the neighborhood of the singular point of the $\mu$ plane. To this aim, we extract the pole contributions from the logarithmic derivative  \eqref{asympt_log} that are closest to $\e=0$:
\begin{equation}\label{airy_poles}
    \partial_\e\ln\Ai(h_-)=\frac{c^{2/3}u^{1/3}}{h_-+r_-}+\cdots,\qquad\partial_\e\ln\Bi(h_+)=\frac{c^{2/3}u^{1/3}}{h_++r_+}+\cdots,
\end{equation}
where $\Ai(-r_-)=\Bi(-r_+)=0$ and
\begin{equation}
    r_-\approx2.34,\qquad r_+\approx1.17.
\end{equation}
Substituting \eqref{airy_poles} into \eqref{integral2}, integrating over $\e$, and stretching the integration variable $u\rightarrow2wu$, we obtain
\begin{equation}\label{integral3}
    c^{\nu-5/2}(2w)^{-l}\int_0^\infty duu^{-1-l}\big[u^2-2(\bar{\mu}\pm w)u+1+r_\mp(2wu/c)^{2/3}\big]^{3/2+l-\nu},
\end{equation}
for the contributions of each of the poles.

The integral \eqref{integral3} is singular when the expression in the square brackets vanishes for some $u=u_0\geq0$. The main contribution to this singularity comes from the vicinity of the point $u=u_0$. Expanding the expression in the square brackets near this point and keeping only the leading terms, we have
\begin{equation}\label{expansion}
\begin{split}
    f(u)&:=u^2-2(\bar{\mu}\pm w)u+1+r_\mp(2wu/c)^{2/3}\approx f(u_0)+\frac12f''(u_0)(u-u_0)^2,\\
    u_0&\approx \bar{\mu}\pm w-\frac{a_\pm}{3}(\bar{\mu}\pm w)^{-1/3},\qquad f(u_0)\approx-\Big(2+\frac{a_\pm}{3}\Big),\qquad \frac12f''(u_0)\approx1-\frac{a_\pm}{9},
\end{split}
\end{equation}
where $a_\pm:=r_\mp(2w/c)^{2/3}$. The integral \eqref{integral3} diverges when $s_\pm:=\bar{\mu}\pm w$ approach the point
\begin{equation}
    s_0(a_\pm)\approx1+a_\pm/2-a^2_\pm/72.
\end{equation}
Substituting the expansion \eqref{expansion} into the integral \eqref{integral3} for $l=-1$ and integrating, we come to
\begin{equation}\label{singl_part}
    2wc^{\nu-5/2}\frac{\sqrt{\pi}\Ga(\nu-1)}{2^{\nu-1}\Ga(\nu-1/2)}(s_0(a_\pm)-s_\pm)^{1-\nu}=2wc^{\nu-5/2}(s_0(a_\pm)-s_\pm)\Big(\nu^{-1}-\ln\frac{s_0(a_\pm)-s_\pm}{2e^{-1}}+O(\nu)\Big),
\end{equation}
in the leading order in $(w/c)$. Keeping only the singular term at $s_\pm\rightarrow s_0(a_\pm)$ and replacing $s_0(a_\pm)\rightarrow1$, which is justified in the leading order in $(w/c)$, we deduce
\begin{equation}\label{sigma_1}
\begin{split}
    \s^{-1}_\nu(\mu)\approx\,& e^{i\pi\nu}\frac{m^{3-2\nu}}{12\pi}S\int_{-L/2}^{L/2}dz\Big\{(1-\tilde{\mu}^2)^{3/2}+\frac{3\tilde{\mu}}{2\pi}\Ga(\nu-1)+\frac{\tilde{\mu}^3}{\pi}\Ga(\nu)+\\
    &+\frac2{\pi} \Big[(1-\tilde{\mu}^2)^{3/2}\arcsin\tilde{\mu}+\tilde{\mu}\Big(\frac43\tilde{\mu}^2-1\Big) \Big] \Big\}+\\
    &+\frac{m^2S}{8\pi^2}\big[ (1-\bar{\mu}-w)\ln(1-\bar{\mu}-w) +(1-\bar{\mu}+w)\ln(1-\bar{\mu}+w) \big]
    -\frac{m^2S}{16\pi}\frac{\bar{\mu}^2+w^2-1}{\nu},
\end{split}
\end{equation}
where, in the last two terms, it was taken into account that $\nu\rightarrow0$. The pole at $\nu\rightarrow0$ in the expression \eqref{singl_part} was discarded as its contribution is taken into account in the last term in \eqref{sigma_1}. The finite terms at $\nu\rightarrow0$ in \eqref{singl_part}, which are not singular for $s\rightarrow s_0$, can also be disregarded to the accuracy of the approximations made.

For $w/c\ll1$, the expression \eqref{sigma_1} gives a good approximation for $\s^{-1}_\nu(\mu)$ and its derivatives with respect to the chemical potential on the interval $|\tilde{\mu}|<1$. The expression \eqref{sigma_1} has a branch point at $\tilde{\mu}=1$ and becomes complex for $\tilde{\mu}>1$. However, the exact expression for $\s^{-1}(\mu)$ is real-valued and well-defined for $s_\pm<s_0(a_\pm)$. That is, there exists the interval of values of the chemical potential, which shrinks to zero for $w/c\rightarrow0$,
\begin{equation}
    1<\bar{\mu}+w<s_0(a_+),\qquad 1<\bar{\mu}-w<s_0(a_-),
\end{equation}
where the expression \eqref{sigma_1} is not applicable. One can improve the expansion \eqref{asympt_log} such that, having integrated over $\e$ and $u$, the leading contribution will approximate $\s^{-1}_\nu(\mu)$ uniformly on the whole interval of physically acceptable values of $\mu$. To this end, one needs to put
\begin{equation}
    h_\pm=(h_\pm+r_\mp)-r_\mp
\end{equation}
in the arguments of the Airy functions and employ the asymptotic expansion in \eqref{asympt_log} supposing that the expression in the parenthesis is large. Then, on evaluating the integral over $\e$, the integral \eqref{hypergeom_int} is replaced by
\begin{equation}
    c^{\nu-5/2}(2w)^{-l-\alpha}\int_0^\infty duu^{-1-l-\alpha}(u^2-2s_\pm u+1+a_\pm u^{2/3})^{3/2+l-\nu+\alpha},
\end{equation}
which is just the Mellin transform. In the present paper we will not investigate the expression resulting from this procedure.

Let us point out some other properties of \eqref{sigma_1}. The term with logarithmic singularity in \eqref{sigma_1} gives the leading contribution to the total charge and the average number of particles at $s_\pm\rightarrow 1$ in spite of the fact that this contribution is suppressed by the factor $w/c$ in comparison with the first term. The first term in \eqref{sigma_1} can be obtained from $\s^{-1}_\nu(\mu)$ for a free particle \cite{HabWeld81}. One just needs to replace $\bar{\mu}\rightarrow\tilde{\mu}$ and integrate the resulting expression over $z$ (see \cite{ElmfSkag95}). As we have already noted above, the contribution of antiparticles is obtained by replacement  $\mu\rightarrow-\mu$ in the corresponding expression for particles. The contribution of antiparticles to the thermodynamic potential cancels out all the terms in \eqref{sigma_1} that are odd with respect to $\tilde{\mu}$.

\section{High-temperature expansion}

\subsection{Cancellation of singularities}

The expression in square brackets in formula \eqref{expan_bos} is an entire function of the parameter $\nu$, see \cite{KalKaz2}. This fact can be exploited for the indirect verification of the expressions for $\sigma^l_{\nu}$ whose explicit form is known for $l=\overline{0,3}$. In the expansion \eqref{expan_bos} the terms with singularities at  $\nu\rightarrow0$ are
\begin{equation}\label{cancel}
\begin{split}
\beta^0:\quad&\Gamma(-2\nu)\zeta(-2\nu)\zeta_4(\nu)\beta^{2\nu}-\frac{1}{2}\sigma^0_{\nu}\\
\beta^1:\quad&\Gamma(-2\nu)\zeta(-1-2\nu)\zeta_4(\nu)\mu\beta^{2\nu}+\frac{1}{12}\sigma^1_{\nu}(\mu)\\
\beta^3:\quad&\Gamma(-2\nu)\zeta(-3-2\nu)\zeta_4(\nu)\frac{\mu^3}{6}\beta^{2\nu}+\Gamma(-2-2\nu)\zeta(-3-2\nu)\zeta_6(\nu)\mu\beta^{2\nu}-\frac{1}{720}\sigma^3_{\nu}(\mu).
\end{split}
\end{equation}
It has been taken into account that there are no singularities in the first and in the second terms at $\beta^2$ and also that $\zeta_{5,7}(0)=0$, see \eqref{zeta_k_general}. It is easy to convince oneself that all the singularities are canceled in this case. This serves as an indirect verification of the fact that the divergent at $\nu\rightarrow0$ (volume) terms in $\sigma^l_{\nu}(\mu)$ were calculated correctly, e.g., the term with $\Gamma(\nu-2)$ in $\sigma^0_{\nu}$.

Let us extract the terms with singularities at $\nu\rightarrow1/2$:
\begin{equation}
\begin{split}
\beta^0:\quad&\Gamma(1-2\nu)\zeta(1-2\nu)\zeta_3(\nu)\beta^{2\nu-1}-\frac{1}{2}\sigma^0_{\nu}\\
\beta^1:\quad&\Gamma(1-2\nu)\zeta(-2\nu)\zeta_3(\nu)\mu\beta^{2\nu-1}+\frac{1}{12}\sigma^1_{\nu}(\mu)\\
\beta^3:\quad&\Gamma(1-2\nu)\zeta(-2-2\nu)\zeta_3(\nu)\frac{\mu^3}{6}\beta^{2\nu-1}+\Gamma(-1-2\nu)\zeta(-2-2\nu)\zeta_5(\nu)\mu\beta^{2\nu-1}-\frac{1}{720}\sigma^3_{\nu}(\mu).
\end{split}
\end{equation}
It has been taken into account that $\zeta_{4,6}(1/2)=0$, see \eqref{zeta_k_general}.
It is not difficult to check that all the singularities are canceled as well. This verifies that the divergent at $\nu\rightarrow1/2$ (surface) terms in $\sigma^l_{\nu}(\mu)$ were calculated correctly, e.g., the term with $\Gamma(\nu-3/2)$ in $\sigma^0_{\nu}$.

\subsection{Grand thermodynamic potential}
Now we have everything to write down the explicit expression for the high-temperature expansion of the one-loop $\Omega$-potential. Formulas \eqref{zeta_k}, \eqref{sigma_expl}, \eqref{sigma_3}, and \eqref{sigma_1} suffice to obtain the expansion up to $\be^3$ terms. We present here only the expression for the divergent and finite terms as  $\be\rightarrow0$. Thus, for bosons
\begin{equation}\label{omega_b_expl}
\begin{split}
    -\Omega_b=\,&\frac{\pi^2}{90}LS\be^{-4}+S\Big[m\frac{\zeta(3)}{\pi^2}\int_{-L/2}^{L/2}dz\tilde{\mu}-\frac{\zeta(3)}{4\pi}\Big]\be^{-3} +S\Big[m^2\int_{-L/2}^{L/2}dz(2\tilde{\mu}^2-1)-\pi\mu \Big]\frac{\be^{-2}}{24}+\\
    &+\frac{m^3}{12\pi\be}S \int_{-L/2}^{L/2}dz\Big\{(1-\tilde{\mu}^2)^{3/2} -\frac{\tilde{\mu}}{2\pi}(2\tilde{\mu}^2-3)\ln\frac{\be^2m^2}{4e} +\frac2{\pi} \Big[(1-\tilde{\mu}^2)^{3/2}\arcsin\tilde{\mu}+\tilde{\mu}\Big(\frac43\tilde{\mu}^2-1\Big) \Big] \Big\}+\\
    &+\frac{m^2}{8\pi^2\be}S \big[ (1-\bar{\mu}-w)\ln(1-\bar{\mu}-w) +(1-\bar{\mu}+w)\ln(1-\bar{\mu}+w) \big]+\\
    &+\frac{S}{8\pi\beta}\ln\beta e^{-\gamma/2}\Big[\mu^2-m^2+\frac{1}{4}(EL)^2\Big]-\frac{S}{8\pi\beta}(\mu^2+(EL)^2)\\
    &+S\Big[\int_{-L/2}^{L/2}dz\Big\{\frac{m^4}{64\pi^2}\ln\frac{\be^2m^2e^{2\ga-3/2}}{16\pi^2} -\frac{5E^2}{96\pi^2} +\frac{m^4}{16\pi^2}\Big(\tilde{\mu}^2-\frac{\tilde{\mu}^4}{3}\Big) -\frac{m^4}{12\pi^2}T^0_{3/2}(2mL) \Big\}+\\
    &+\frac{m^3}{24\pi}+\frac{\mu(4\mu^2+3(EL)^2-12m^2)}{192\pi}\Big].
\end{split}
\end{equation}
If we take into account the contribution from antiparticles, then
\begin{equation}\label{omega_b_plus_antipart}
\begin{split}
    -\Omega_b=\,&\frac{\pi^2}{45\be^4}LS-\frac{\zeta(3)}{2\pi\be^3}S +\frac{m^2}{12\be^{2}}S\int_{-L/2}^{L/2}dz(2\tilde{\mu}^2-1)
    +\frac{m^3}{6\pi\be}S \int_{-L/2}^{L/2}dz(1-\tilde{\mu}^2)^{3/2}+\\
    &+\frac{m^2}{8\pi^2\be}S \big[ (1-\bar{\mu}-w)\ln(1-\bar{\mu}-w) +(1-\bar{\mu}+w)\ln(1-\bar{\mu}+w)+ \\
    &+(1+\bar{\mu}-w)\ln(1+\bar{\mu}-w) +(1+\bar{\mu}+w)\ln(1+\bar{\mu}+w)\big]+\\
    &+\frac{S}{4\pi\beta}\ln\beta e^{-\gamma/2}\Big[\mu^2-m^2+\frac{1}{4}(EL)^2\Big]-\frac{S}{4\pi\beta}(\mu^2+(EL)^2)\\
    &+S\Big[\int_{-L/2}^{L/2}dz\Big\{\frac{m^4}{32\pi^2}\ln\frac{\be^2m^2e^{2\ga-3/2}}{16\pi^2} -\frac{5E^2}{48\pi^2} +\frac{m^4}{8\pi^2}\Big(\tilde{\mu}^2-\frac{\tilde{\mu}^4}{3}\Big) -\frac{m^4}{6\pi^2}T^0_{3/2}(2mL) \Big\}+\frac{m^3}{12\pi}\Big].
\end{split}
\end{equation}
The expression for the terms at positive powers of $\be$ of the high-temperature expansion (of the one-loop $\Omega$-potential) with the contribution from antiparticles reads as

\begin{equation}\label{posit_powers}
\lim_{\nu\rightarrow0}\Big[\sum^{\infty}_{\substack{k,s=0\\k+2s>4}}\Gamma(4-2\nu-k)\zeta(4-2\nu-k-2s)\zeta_k(\nu)\beta^{2s+k-4}\frac{2\mu^{2s}}{(2s)!}\Big].
\end{equation}
It is easy to see that the derived expression is regular in the $\nu$-plane; if there are singularities in gamma functions, they are annihilated by zeros of the zeta function or the coefficients $\zeta_k(\nu)$, see \eqref{zeta_k}. The vacuum contribution is not taken into account in \eqref{omega_b_expl}, \eqref{omega_b_plus_antipart}, and \eqref{posit_powers}.

The derived formulas can be applied to the systems in a local thermodynamical equilibrium. Namely, let the system be represented as a collection of subsystems with characteristic size $L$, which are large enough for the statistic description to apply, and sufficiently small as the electromagnetic field to be approximated as crossed, constant, and homogeneous on a scale $L$ in a comoving reference frame. It is also assumed that the subsystems possess small acceleration in this frame so as there is no influence of the metric on thermodynamic properties of the subsystem. Then, for such a system, we find from \eqref{omega_b_expl} that
\begin{equation}\label{omega_b_distr}
\begin{split}
    -\Omega_b=\,&\int d^3x\Big[ \frac{\pi^2T^4}{90}+\frac{\zeta(3)T^3}{\pi^2} A_0 -\frac{m^2-2A_0^2}{24}T^2+\\
    &+\frac{T}{12\pi}\Big\{(m^2-A_0^2)^{3/2} -\frac{A_0}{2\pi}(2A_0^2-3m^2)\ln\frac{\be^2m^2}{4e} +\frac2{\pi} \Big[(m^2-A_0^2)^{3/2}\arcsin \frac{A_0}{m}+A_0\Big(\frac43A_0^2-m^2\Big) \Big] \Big\}+\\
    &+\frac{mT}{4\pi^2 L} (m-A_0)\ln\frac{m-A_0}{m}
    +\frac{m^4}{64\pi^2}\ln\frac{m^2e^{2\ga-3/2}}{16\pi^2T^2} -\frac{5E^2}{96\pi^2} +\frac{1}{16\pi^2}\Big(m^2A_0^2-\frac{A_0^4}{3}\Big) \Big],
\end{split}
\end{equation}
in the leading order of $w/c$. Here, $T:=\be^{-1}$, the chemical potential is included into the definition of $A_0:=u^\mu A_\mu$, where $u^\mu$ is a velocity $4$-vector of a medium, and the Coulomb gauge is assumed. The parameter $L$ entering into \eqref{omega_b_distr} and, consequently, the whole coefficient at the term in question have only an order-of-magnitude estimate. Taking into account the contribution from antiparticles, we deduce
\begin{multline}
\begin{split}
    -\Omega_b=\,&\int d^3x\Big[ \frac{\pi^2T^4}{45}-\frac{m^2-2A_0^2}{12}T^2+\frac{T}{6\pi}(m^2-A_0^2)^{3/2}+\frac{mT}{4\pi^2 L} \Big[(m-A_0)\ln\frac{m-A_0}{m}+\\
    &+(m+A_0)\ln\frac{m+A_0}{m}\Big]+\frac{m^4}{32\pi^2}\ln\frac{m^2e^{2\ga-3/2}}{16\pi^2T^2}-\frac{5E^2}{48\pi^2}+\frac{1}{8\pi^2}\Big(m^2A_0^2-\frac{A_0^4}{3}\Big) \Big].
\end{split}
\end{multline}

\subsection{Vacuum energy}

In the paper \cite{KalKaz4} the high-temperature expansion of the grand thermodynamic potential for particles obeying the Fermi-Dirac statistics was derived
\begin{equation}\label{expan_ferm}
    -\Omega_{f}(\mu)\simeq\sum_{k,n=0}^\infty\Gamma(D-2\nu-k)\eta(D-2\nu-k-n)\frac{\zeta_{k}(\nu)(\beta\mu)^n}{n!\beta^{D-2\nu-k}}
    +\sum^{\infty}_{l=0}\frac{(-1)^l \eta(-l)}{\Gamma(l+1)}\sigma^l_{\nu}(\mu)\beta^l,\quad \nu\rightarrow0.
\end{equation}
Here, in contrast to formula \eqref{expan_bos}, all the Riemann zeta functions are replaced by the Dirichlet eta functions $\eta(z):=(1-2^{1-z})\zeta(z)$ and, additionally, the term with $l=-1$ is absent because the eta function has no singularity at $z=1$.

There is an interesting fact that the high-temperature expansion for fermions \eqref{expan_ferm} can be used to find the contribution to the effective action at zero temperature (vacuum energy). It rests on the following observation
\begin{equation}\label{E_vac}
\partial_{\beta_0}(\beta_0\Omega_f(\mu=0,\beta_0))=\sum_{n}\frac{E_n}{e^{\beta_0 E_n}+1}\underset{\be_0\rightarrow0}{\rightarrow}\sum_{n}\frac{E_n}{2}=E_{vac},
\end{equation}
where $E_n$ is the energy of the mode with number $n$. It should be noted that the Fermi-Dirac distribution in this formula plays the role of a regularizing factor, and $\beta_0$ is a regularization parameter. The derived formula gives the unrenormalized energy of vacuum fluctuations for one bosonic degree of freedom.

To obtain the vacuum energy, one can use the high-temperature fermionic expansion:
\begin{multline}
    \Omega_f\big|_{\beta_0\rightarrow0}=-\frac{7\pi^2}{720}V\beta_0^{-4}+\frac{3\zeta(3)}{16\pi}S\beta_0^{-3}+\frac{1}{48}\Big[m^2-\frac{1}{6}(EL)^2\Big]V\beta_0^{-2}-\frac{\ln 2}{8\pi}\Big[m^2-\frac{1}{4}(EL)^2\Big]S\beta_0^{-1}-\\
    -\frac{V}{3840\pi^2}\Big[(EL)^4+90m^4-20(EL)^2m^2+200E^2\Big]+\frac{m^3}{24\pi}S+\frac{m^4 V}{32\pi^2}\ln\frac{m\beta_0 e^{\gamma}}{\pi}-\frac{m^4 V}{12\pi^2}T^{0}_{3/2}(2mL)+\cdots
\end{multline}
Then for the vacuum energy we have
\begin{multline}
    E_{vac}=2\partial_{\beta_0}(\beta_0 \Omega_f(0))=\frac{7\pi^2}{120}V\beta_0^{-4}-\frac{3\zeta(3)}{4\pi}S\beta_0^{-3}-\frac{1}{24}\Big[m^2-\frac{1}{6}(EL)^2\Big]V\beta_0^{-2}+\frac{m^4 V}{16\pi^2}\ln\frac{m\beta_0 e^{\gamma}}{\pi}-\\
    -\frac{V}{1920\pi^2}\Big[(EL)^4-30m^4-20m^2(EL)^2+200E^2\Big]+\frac{m^3}{12\pi}S-\frac{m^4 V}{6\pi^2}T^0_{3/2}(2mL).
\end{multline}
Introduce the following counter-terms
\begin{multline}
    c.t.=\frac{7\pi^2}{120}\beta_0^{-4}-\frac{3\zeta(3)}{4\pi}\frac{1}{L}\beta_0^{-3}-\frac{1}{24}\Big[m^2-\frac{1}{6}(EL)^2\Big]\beta_0^{-2}+\frac{m^4}{16\pi^2}\ln\frac{m\beta_0 e^{\gamma}}{\pi}-\\
    -\frac{1}{1920\pi^2}\Big[(EL)^4-30m^4-20m^2(EL)^2+200E^2\Big]+\frac{m^3}{12\pi}\frac{1}{L}.
\end{multline}
Consequently, the renormalized vacuum energy reads as
\begin{equation}
    \frac{E^{ren}_{vac}}{V}=\frac{E_{vac}}{V}-c.t.=-\frac{m^4}{6\pi^2}T^0_{3/2}(2mL),
\end{equation}
which is twice as big as (antiparticle contribution included) the renormalized vacuum energy for a massive scalar field with the Dirichlet boundary conditions in the absence of any electromagnetic fields (see, e.g., \cite{Khusnut2,Ambjorn,BordKlim}).

\section{Conclusion}

Let us sum up the results. We obtained the explicit expressions for the high-temperature expansion of the one-loop contribution to the thermodynamic potential of charged scalar particles in a constant crossed electromagnetic field. The vacuum energy was also calculated. The contributions of particles and antiparticles were separately investigated. It was shown explicitly that the high-temperature expansion of the thermodynamic potential and the vacuum energy do not contain contributions that are exponentially suppressed with respect to the external field or coupling constant except, possibly, the term at $1/\be$.

The fact that the non-perturbative corrections to the vacuum energy do not depend on the external field for the given configuration of fields and plates can be anticipated from a simple analysis of the gauge invariant Lorentz-invariants. For the crossed electromagnetic field and given boundary conditions, the stress tensor has the form $F_{\mu\nu}:=E^2h_{[\mu}n_{\nu]}$,
where
\begin{equation}
    h^{\mu}=(1,1,0,0),\quad n^{\mu}=(0,0,0,1)\quad\Rightarrow\quad hn=0,\quad n^2=-1,\quad h^2=0.
\end{equation}
Here $n^\mu$ is the normal to the hypersurface $z=\pm L/2$. Due to antisymmetry of $F_{\mu\nu}$, the following scalars vanish: $\eta^{\mu\nu}F_{\mu\nu}=n^{\mu}n^{\nu}F_{\mu\nu}=0$. All the higher powers of $F_{\mu\nu}$ are also zero, $F^k_{\mu\nu}=0$, apart from $F^2_{\mu\nu}=E^2 h_{\mu}h_{\nu}$. Nevertheless, $\eta^{\mu\nu}F^2_{\mu\nu}=n^{\mu}n^{\nu}F^2_{\mu\nu}=0$. The same considerations apply to the dual tensor $\tilde{F}_{\mu\nu}=\frac{1}{2}\varepsilon_{\mu\nu\alpha\beta}F^{\alpha\beta}$, $\tilde{F}^2_{\mu\nu}=E^2 h_{\mu}h_{\nu}$. The cross products also vanish, $(\tilde{F}F)_{\mu\nu}=0$. Notice that the presence of non-perturbative contributions in $\sigma^{-1}_{\nu}$ that depend on the external field is not ruled out. These contributions may depend on the $4$-vector specifying the reference frame where the thermodynamic system is at rest. Besides, in deriving the explicit expression for the thermodynamic potential, we essentially rely on the fulfillment of the condition $EL<2m$ in the comoving reference frame. When this condition is violated, the additional contribution to the effective action can arise.

The system considered in the present paper can be used for an approximate description of thermodynamical properties of an ultrarelativistic fluid of charged bosons under the assumption that it is in a local thermodynamic equilibrium. In the reference frame comoving with the fluid element, the external electromagnetic field is crossed with good accuracy. Therefore, if the acceleration of this fluid element is small such that the effect of inertial forces on the thermodynamic properties of the system is negligible, then the thermodynamic properties of such a fluid element can be described by formulas obtained in the present paper.

%\newpage
\appendix

\section{Dyson series}\label{Dyson}

A heat kernel (the exponent on the right-hand side of \eqref{zeta_+}) can be considered as an evolution operator specified by the Hamiltonian $-H(\omega)$ \eqref{KG_spectrum} taken at imaginary time. The dependence of the evolution operator on the constant $p_x^2+p_y^2+m^2-\omega^2$ is trivial and will be taken into account only in the final answer. We use the perturbation theory (see, e.g., \cite{KalKaz2}) to find the matrix elements of the evolution operator taking $H_0=\hat{p}^2_z$ as an unperturbed Hamiltonian and $-2(\omega+p_x)E\hat{z}$ as a perturbation.

The solution to the Sturm-Liouville problem for the Hamiltonian $H_0$ with zero boundary conditions \eqref{KG_spectrum} is given by the following functions
\begin{equation}
    \psi^0_k(z)=\sqrt{\frac{2}{L}}\sin\big[\sqrt{\lambda_k}(z+\frac{L}{2})\big],\qquad \lambda_k=\frac{\pi^2 k^2}{L^2},\quad k=\overline{1,\infty},
\end{equation}
It is easy to see that in the basis formed by these functions the operator of coordinate and the unperturbed Hamiltonian have the following matrix elements:
\begin{equation}
    H^0_{nk}=\frac{\pi^2 n^2}{L^2}\delta_{nk},\qquad z_{nk}=\frac{4L}{\pi^2}nk \frac{(-1)^{n+k}-1}{(n^2-k^2)^2}\quad\text{if}\quad n\neq k,\qquad z_{nk}=0\quad \text{if}\quad n=k.
\end{equation}
It should be noted that despite the fact that $H^0$ is quadratic in momentum $[x,[x,[x,H^0]]]\neq 0$. This is the reason why the naive expression for the heat kernel in external homogeneous electromagnetic fields \cite{KalKazIzv,ElmfSkag95} is not applicable in our case. Time evolution operator takes an especially simple form in the interaction picture (Dyson series)
\begin{equation}
    e^{-i \hat{H} s}=1+2i(\omega+p_x) E \int_{0}^{s}\hat{z}(t_1)d t_1-4(\omega+p_x)^2 E^2 \int_{0}^{s}\hat{z}(t_1)d t_1 \int_{0}^{t_1}\hat{z}(t_2) d t_2+\ldots,
\end{equation}
where  $\hat{z}(t)=e^{itH_0}\hat{z}e^{-itH_0}$ is the coordinate operator in the interaction representation.

Consider the correction of the type $E^0$:
\begin{equation}
    \sum^{\infty}_{n=1}\langle n|e^{-is\hat{H}}|n\rangle^{(0)}\approx\sum^{\infty}_{n=1}e^{-is\frac{\pi^2 n^2}{L^2}}
\end{equation}
Taking into account the constant term discarded in $-H(\omega)$ and changing $s=i\tau$, we obtain the zeroth approximation for the heat kernel
\begin{equation}
    G^{(0)}(\omega,\tau;p_x,p_y)=e^{\tau(p_x^2+p_y^2+m^2-\omega^2)}\sum^{\infty}_{n=1} e^{\tau\frac{\pi^2 n^2}{L^2}},
\end{equation}
where we set $\re \tau<0$ for the sake of convergence.

There is a trace of the heat kernel in the expression for the high-temperature expansion
\begin{equation}
    \Sp G(\omega,\tau)=S \int \frac{dp_xdp_y}{(2\pi)^2} G(\omega,\tau;p_x,p_y),
\end{equation}
where $S:=L_xL_y$. The integrals over momenta are Gaussian and can easily be evaluated
\begin{equation}
    \Sp G^{(0)}(\omega,\tau)=-\frac{S}{4\pi\tau}e^{-\tau(\omega^2-m^2)}\sum_{n=1}^{\infty}e^{\tau\frac{\pi^2 n^2}{L^2}}.
\end{equation}
It is not difficult to find an asymptotic expansion of the remaining sum at small values of $\tau$:
\begin{equation}
    \sum_{n=1}^{\infty}e^{\tau\frac{\pi^2 n^2}{L^2}}\approx\frac{1}{2}\frac{L}{\pi^{1/2}}(-\tau)^{-1/2}-\frac{1}{2}.
\end{equation}

In order to calculate the expansion of $\zeta_+(\nu,\omega)$ at $\omega\rightarrow\infty$, it is convenient to use the following relation
\begin{equation}
    \zeta_+(\nu,\omega)\approx \int_{C} \frac{d\tau}{2\pi i}\tau_{+}^{\nu-1}\Big(\sum_{k}a_{k/2}(-\tau)_{-}^{k/2}\Big)e^{-\tau(\omega^2-m^2)} =e^{i\pi\nu}\sum_{k}\frac{a_{k/2}(\omega^2-m^2)^{-\nu-k/2}}{\Gamma(1-\nu-k/2)},
\end{equation}
where we have already taken into account the connection between different branches of the square root function characterized by cuts along the negative $(-)$ and positive $(+)$ axes: $(-\tau)^{k/2}_{-}=e^{-i\frac{\pi}{2}k} \tau_{+}^{k/2}$.

Thus, for the $E^0$ correction we deduce
\begin{equation}
    \zeta^{(0)}_+(\nu,\omega)\approx\frac{e^{i\pi\nu}V}{8\pi^{3/2}} \frac{(\omega^2-m^2)^{3/2-\nu}}{\Gamma(5/2-\nu)}-\frac{e^{i\pi\nu}S}{8\pi}\frac{(\omega^2-m^2)^{1-\nu}}{\Gamma(2-\nu)}.
\end{equation}
This gives the following contribution into the coefficients $\zeta_k$:
\begin{equation}
\begin{split}
    \zeta_0&\approx \frac{e^{i\pi\nu}V}{8\pi^{3/2}}\frac{1}{\Gamma(5/2-\nu)},\quad
    \zeta_1\approx -\frac{e^{i\pi\nu}S}{8\pi}\frac{1}{\Gamma(2-\nu)},\quad
    \zeta_2\approx -\frac{e^{i\pi\nu}V}{8\pi^{3/2}}\frac{m^2}{\Gamma(3/2-\nu)},\quad
    \zeta_3\approx \frac{e^{i\pi\nu}S}{8\pi}\frac{m^2}{\Gamma(1-\nu)},\\
    \zeta_4&\approx \frac{e^{i\pi\nu}V}{8\pi^{3/2}}\frac{m^4}{2\Gamma(1/2-\nu)},\quad
    \zeta_5\approx -\frac{e^{i\pi\nu}S}{8\pi}\frac{m^4}{2\Gamma(-\nu)},\quad
    \zeta_6\approx -\frac{e^{i\pi\nu}V}{8\pi^{3/2}}\frac{m^6}{6\Gamma(-1/2-\nu)}.
\end{split}
\end{equation}
There is no correction of the form $E^1$ since $z_{nn}=0$. Consider the correction of the form $E^2$ (denote it by index $(2)$):
\begin{equation}
    \sum^{\infty}_{n=1}\langle n|e^{-is\hat{H}}|n\rangle_{(2)}\approx-4(\omega+p_x)^2 E^2 \sum^{\infty}_{n,p=1}is\frac{L^2}{\pi^2}\frac{z^2_{np}}{n^2-p^2}e^{-is\frac{\pi^2 n^2}{L^2}}.
\end{equation}
The sum over $p$ is readily evaluated and the contribution to the heat kernel reads as
\begin{equation}
    G^{(2)}(\omega,\tau;p_x,p_y)=e^{\tau(p_x^2+p_y^2+m^2-\omega^2)}(\omega+p_x)^2 E^2 \sum^{\infty}_{n=1}\tau\frac{L^4}{12\pi^4}\Big(\frac{\pi^2}{n^2}-\frac{15}{n^4}\Big)e^{\tau\frac{\pi^2 n^2}{L^2}}.
\end{equation}
Having calculated the trace in the phase space, we arrive at
\begin{equation}
    \Sp G^{(2)}(\omega,\tau)=\frac{VE^2L^3}{48\pi^5}\Big(\frac{1}{2\tau}-\omega^2\Big)e^{-\tau(\omega^2-m^2)}\sum_{n=1}^{\infty}\Big(\frac{\pi^2}{n^2}-\frac{15}{n^4}\Big)e^{\tau\frac{\pi^2 n^2}{L^2}}.
\end{equation}
The remaining sum over $n$ has the following asymptotic expansion in small $\tau$:
\begin{equation}
\sum_{n=1}^{\infty}\Big(\frac{\pi^2}{n^2}-\frac{15}{n^4}\Big)e^{\tau\frac{\pi^2 n^2}{L^2}}\approx-\frac{\pi^{7/2}}{L}\Big( (-\tau)^{1/2}+3\frac{\pi^{1/2}}{L}(-\tau)+10\frac{1}{L^2}(-\tau)^{3/2}-\frac{15}{4}\frac{\pi^{1/2}}{L^3}(-\tau)^{2}\Big).
\end{equation}
This gives the following contribution into the coefficients $\zeta_k$:
\begin{equation}
\begin{split}
    \zeta_2&\approx -\frac{e^{i\pi\nu}V}{8\pi^{3/2}}\frac{E^2L^2(\nu-1)}{6\Gamma(3/2-\nu)},\quad
    \zeta_3\approx \frac{e^{i\pi\nu}S}{8\pi}\frac{E^2 L^2(\nu-1/2)}{2\Gamma(1-\nu)},\quad
    \zeta_4\approx \frac{e^{i\pi\nu}V}{8\pi^{3/2}}\frac{\nu E^2(L^2 m^2-10)}{6\Gamma(1/2-\nu)},\\
    \zeta_5&\approx -\frac{e^{i\pi\nu}S}{8\pi}\frac{(\nu+1/2)E^2(4L^2 m^2-5)}{8\Gamma(-\nu)},\quad
    \zeta_6\approx -\frac{e^{i\pi\nu}V}{8\pi^{3/2}}\frac{E^2 m^2(L^2 m^2-20)(\nu+1)}{12\Gamma(-1/2-\nu)}.
\end{split}
\end{equation}
Collecting together all the above formulas, we arrive at
\begin{equation}
\begin{split}
	\zeta_0(\nu)&=\frac{e^{i\pi\nu}V}{8\pi^{3/2}}\frac{1}{\Gamma(5/2-\nu)},\\
	\zeta_1(\nu)&=-\frac{e^{i\pi\nu}S}{8\pi}\frac{1}{\Gamma(2-\nu)},\\
	\zeta_2(\nu)&=-\frac{e^{i\pi\nu}V}{8\pi^{3/2}}\frac{1}{\Gamma(3/2-\nu)}\Big[m^2+\frac{1}{6}(\nu-1)(EL)^2\Big],\\
	\zeta_3(\nu)&=\frac{e^{i\pi\nu}S}{8\pi}\frac{1}{\Gamma(1-\nu)}\Big[m^2+\frac{1}{2}(\nu-\frac{1}{2})(EL)^2\Big],\\
	\zeta_4(\nu)&=\frac{e^{i\pi\nu}V}{8\pi^{3/2}}\frac{1}{\Gamma(1/2-\nu)}\Big[\frac{1}{2}m^4+\frac{1}{6}\nu(EL)^2m^2-\frac{5}{3}\nu E^2\Big],\\
	 \zeta_5(\nu)&=-\frac{e^{i\pi\nu}S}{8\pi}\frac{1}{\Gamma(-\nu)}\Big[\frac{1}{2}m^4+\frac{1}{2}(\nu+\frac{1}{2})(EL)^2m^2-\frac{5}{8}(\nu+\frac{1}{2})E^2\Big],\\
	 \zeta_6(\nu)&=-\frac{e^{i\pi\nu}V}{8\pi^{3/2}}\frac{1}{\Gamma(-1/2-\nu)}\Big[\frac{1}{6}m^6+\frac{1}{12}(\nu+1)(EL)^2m^4-\frac{5}{3}(\nu+1)E^2 m^2\Big].
\end{split}
\end{equation}
The corrections of the form $E^4$ and $E^6$ can be found analogously. It is convenient to perform the calculation by using a computer. Of course, the answer coincides with formula \eqref{zeta_k}.

%\newpage
\section{Some series}\label{Some_Series}

Let us consider the series (the Epstein-Hurwitz zeta function, see \cite{Khusnut1,ChowSelb})
\begin{equation}\label{series_0}
    F^0_\al(c):=\sum_{n=1}^\infty\Big(1+\frac{\pi^2n^2}{4c^2}\Big)^\al,\qquad\re\al<-1/2,
\end{equation}
that appears in the expression for $\sigma^0_\nu(\mu)$ in \eqref{sigma_expl}. This series is a holomorphic function in the indicated domain of the complex $\al$ plane and can be continued by analyticity to the whole complex plane except a countable number of poles. Applying the Poisson summation formula to the series of the form \eqref{series_0}, but with infinite limits, it is not difficult to derive
\begin{equation}
    F^0_\al(c)=-\frac12+\frac{c}{\sqrt{\pi}}\frac{\Ga(-\al-1/2)}{\Ga(-\al)}-\frac{2c}{\pi}\sin(\pi\al)T^0_\al(4c),
\end{equation}
where
\begin{equation}\label{T0}
\begin{split}
    T^0_\al(\be):&=\int_1^\infty dx\frac{(x^2-1)^\al}{e^{\be x}-1}=\sum_{p=1}^\infty\frac{\Ga(\al+1)}{\sqrt{\pi}}\frac{K_{\al+1/2}(p\be)}{(p\be/2)^{\al+1/2}}=\\ &=\frac{\Ga(\al+1)}{8\pi^{3/2}i}\int_C ds\Ga\Big(-\al-\frac{s+1}{2}\Big)\Ga\Big(-\frac{s}{2}\Big)\zeta(-s)\Big(\frac{\be}{2}\Big)^s,\qquad \re\al>-1,
\end{split}
\end{equation}
where the contour $C$ runs upward parallel to the imaginary axis such that $\re(s+2\al)<-1$ and $\re s<-1$. The last two expressions in \eqref{T0} can be used to construct the analytic continuation with respect to $\al$ to the region $\re\al\leq-1$. The last formula in \eqref{T0} follows from the Mellin representation (see, e.g., \cite{Davies}):
\begin{equation}\label{BE_Mellin}
    (e^x-1)^{-1}=\int_{C}\frac{ds}{2\pi i}\Ga(-s)\zeta(-s)x^s,
\end{equation}
where the contour $C$ goes upward from below parallel to the imaginary axis and to the left from the point $s=-1$.

The series,
\begin{equation}
    F^4_\al(c):=\sum_{n=1}^\infty n^{-4}\Big(1+\frac{\pi^2n^2}{4c^2}\Big)^\al,\qquad\re\al<3/2,
\end{equation}
also arises in \eqref{sigma_expl}. It is useful to write it as
\begin{equation}
    F^4_\al(c):=\sum_{n=1}^\infty n^{-4}[(1+xn^2)^\al-1-\al xn^2]+\sum_{n=1}^\infty(n^{-4}+\al xn^{-2}),
\end{equation}
where $x:=\pi^2/(4c^2)$. The latter series is reduced to the sum of the zeta functions. The former series can be completed to the series with infinite limits, the term with $n=0$ being understood as a limit. Then, using the Poisson formula, we deduce
\begin{equation}
    F^4_\al(c)=\zeta(4)+\al\zeta(2)\frac{\pi^2}{4c^2}-\frac{\al(\al-1)}{64}\frac{\pi^4}{c^4}+\frac{\Ga(3/2-\al)}{\Ga(-\al)}\frac{\pi^{7/2}}{12 c^3}-\frac{\pi^3}{4c^3}\sin(\pi\al)T^4_\al(4c),
\end{equation}
where
\begin{equation}\label{T4}
    T^4_\al(\be):=\int_1^\infty \frac{dx}{x^4}\frac{(x^2-1)^\al}{e^{\be x}-1} =\Ga(\al+1)\int_C \frac{ds}{4\pi i}\frac{\Ga(-\al-s/2+3/2)\Ga(-s)\zeta(-s)}{\Ga(5/2-s/2)}\be^s,\qquad \re\al>-1,
\end{equation}
and the contour $C$ runs upward parallel to the imaginary axis such that $\re(s+2\al)<3$ and $\re s<-1$. The contributions \eqref{T0}, \eqref{T4} are exponentially suppressed for $c\rightarrow+\infty$,
\begin{equation}
    T^{2k}_{\alpha}(4c)=\int_1^\infty \frac{dx}{x^{2k}}\frac{(x^2-1)^\al}{e^{4cx}-1}\approx\frac{1}{2}\frac{\Gamma(\alpha+1)}{(2c)^{\alpha+1}}e^{-4c}.
\end{equation}

\begin{acknowledgments}

The work was supported by the RFBR grant 20-32-70023.

\end{acknowledgments}

\end{document}